\documentclass[a4paper,twocolumn,aps,superscriptaddress,floatfix]{revtex4-1}

\usepackage{graphicx}
\usepackage{comment}
\usepackage{amsmath,amsfonts,amssymb,amsthm,stmaryrd}
\usepackage{mathtools,braket}
\usepackage{url}
\usepackage{ulem}
\usepackage{siunitx}
\usepackage{hyperref}
\usepackage{color}
\usepackage[utf8]{inputenc}
\usepackage{makeidx}

\definecolor{darkolivegreen}{rgb}{0.33, 0.42, 0.18}

\definecolor{cadmiumorange}{rgb}{0.93, 0.53, 0.18}
\definecolor{cadmiumgreen}{rgb}{0.0, 0.42, 0.24}

\def\ai	{{\textit{ab initio}}}

\def\({\left(}
\def\){\right)}
\def\[{\left[}
\def\]{\right]}

\def\bfQ{\mathbf{Q}}

\def\bfk{\mathbf{k}}

\def\bfq{\mathbf{q}}
\def\bfr{\mathbf{r}}

\def\bfj{\mathbf{j}}
\def\bfv{\mathbf{v}}

\def\bfP{\mathbf{P}}
\def\bfQ{\mathbf{Q}}
\def\bfE{\mathbf{E}}

\def\bfR{\mathbf{R}}
\def\bfG{\mathbf{G}}
\def\>{\rangle}
\def\<{\langle}

\def\a{\alpha}

\def\G{\Gamma}
\def\d{\delta}

\def\l{\lambda}

\def\vf{\varphi}

\def\x{\chi}
\def\w{\omega}

\def\q{\psi}

\newcommand{\be}{\begin{equation}}
\newcommand{\ee}{\end{equation}}
\newcommand{\ba}{\begin{eqnarray}}
\newcommand{\ea}{\end{eqnarray}}
\newcommand{\nn}{\notag}

\newcommand{\lb}{\label}

\newcommand{\op}[1]{\hat {#1}}

\def\dipole{\op{\boldsymbol\mu}}
\def\opr{\op{\bfr}}
\def\opv{\op{\bfv}}
\def\opR{\op{\bfR}}

\def\efield{\boldsymbol{\mathcal{E}}}

\def\ks{Kohn-Sham}

\newcommand{\ac}[1]{\op{a}^{\dagger}_{{#1}}}
\newcommand{\ad}[1]{\op{a}_{{#1}}}
\newcommand{\psic}[1]{\op{\q}^{\dagger}({#1})}
\newcommand{\psid}[1]{\op{\q}({#1})}

\newcommand{\subalign}[1]{%
  \vcenter{%
    \Let@ \restore@math@cr \default@tag
    \baselineskip\fontdimen10 \scriptfont\tw@
    \advance\baselineskip\fontdimen12 \scriptfont\tw@
    \lineskip\thr@@\fontdimen8 \scriptfont\thr@@
    \lineskiplimit\lineskip
    \ialign{\hfil$\m@th\scriptstyle##$&$\m@th\scriptstyle{}##$\crcr
      #1\crcr
    }%
  }
}

\begin{document}

\author{D. Sangalli}
\affiliation{Istituto di Struttura della Materia-CNR (ISM-CNR), Area della Ricerca di Roma 1, Monterotondo Scalo, Italy}

\author{M. D'Alessandro}
\affiliation{Istituto di Struttura della Materia-CNR (ISM-CNR), Via del Fosso del Cavaliere 100, 00133 Roma, Italia}

\author{C. Attaccalite}
\affiliation{CNRS/Mix-Marseille Universit\'e, Centre Interdisciplinaire de Nanoscience de Marseille UMR 7325 Campus de Luminy, 13288 Marseille Cedex 9, France}

\title{Exciton - Exciton transitions involving strongly bound excitons:\\an \ai\ approach}
\date{\today}

\begin{abstract}
In pump-probe spectroscopy, two laser pulses are employed to garner dynamical information from the sample of interest. The pump initiates the optical process by exciting a portion of the sample from the electronic ground state to an accessible electronic excited state, an exciton. Thereafter, the probe interacts with the already excited sample. The change in the absorbance after pump provides information  on transitions between the excited states and their dynamics.\\
In this work we study these exciton-exciton transitions by means of an \ai\ real time propagation scheme based on dynamical Berry phase formulation. The results are then analyzed taking advantage of a Fermi-golden rule approach formulated in the excitonic basis-set and in terms of the symmetries of the excitonic states. Using bulk LiF and 2D hBN as two prototype materials, we discuss the selection rules for transitions involving strongly bound Frenkel excitons, for which the hydrogen model cannot be used.
\end{abstract}

\maketitle

\section{Introduction}

Excitons are composite particles, formed by bound electron-hole (eh) pairs. Optically bright excitons dominate the equilibrium absorption spectrum of a wide class of materials, and can be easily measured~\cite{toyozawa2003optical}.
While in standard semi-conductors their binding energy is of few meV, in other systems such as 2D materials~\cite{thygesen2017calculating}, insulators like LiF and hBN~\cite{Benedict1998,Rohlfing1998,Rohlfing2000,Kubota2007,galvani2016excitons}, semiconductors like Cu$_2$O~\cite{Kazimierczuk2014} or BiI$_3$~\cite{Kaifu1988,Mor2021}, or in organic semi-conductors~\cite{Bakulin2012,Gelinas2011}, it can be as high as hundreds of meV, making excitons stable at room temperature. As a consequence optically injected excitons can be exploited for opto-electronic devices\cite{mueller2018exciton}. This calls for a detailed understanding of the excitonic properties and their dynamics.

Different phenomena can participate in the exciton dynamics. Excitons can scatter with defects, phonons, annihilate each other (Auger effect) or end up in dark states\cite{selig2019ultrafast,Paleari2019,PhysRevMaterials.7.024006,Chen2020,antonius2022theory,Chen2022}, and
the knowledge of these state is crucial to predict the dynamics. However, while bright excitons can be easily investigated both experimentally and theoretically, dark excitons are much more difficult to measure. The validation of numerical modeling can be often achieved only in a very indirect way, via accurate comparison between numerical simulations and experimental measurements.
Time-resolved angle-resolved photo-emission (TR-ARPES) measurements can be used to measure both dark and bright excitons~\cite{Perfetto2019b,Sangalli2021}. However, TR-ARPES experiments require complex experimental setup, often have limited energy resolution and are restrained to the study of the lowest energy excitons~\cite{Dong2021,Michael2021}. 
Two possible setup to directly investigate dark excitons via table-top absorption experimental setup are nonlinear optics experiments at equilibrium, and transient absorption (TR-abs) experiments in the non-equilibrium regime.
In the non-linear regime, two photon absorption have been used to directly excite dark excitons in 2D~\cite{cassabois2016hexagonal,Ye2014,Panna2019} and 1D\cite{wang2005optical} materials, and accurate schemes based on \ai\ simulations have been developed to describe these experiments~\cite{Attaccalite2018}.
In the non-equilibrium regime, dark excitons can be explored via TR-Abs by combining optical and THz (or infrared) laser pulses.
A first optical pump pulse is used, with frequency tuned resonant to some excitonic peak, and a second THz probe pulse is used to measure transitions from the initially injected bright exciton towards available dark excitons.
This was explored for example in GaAs quantum wells~\cite{Huber2005,Kaindl2003}, in bulk silicon~\cite{Suzuki2009,Suzuki2012} and in materials with larger excitonic binding energy, such as Cu$_2$O~\cite{Jorger2005}, 2D layers~\cite{Poellmann2015,Cha2016,Steinleitner2018,Merkl2019}, and hybrid organic-inorganic semiconductors~\cite{Luo2017}. However, no \ai\, approach has been so far developed to model these experiments.

Exciton to exciton transitions are usually divided into two groups: (i) inter-exciton transitions, if the initial and final state belong to two different excitonic series or (ii) intra-exciton transitions, if the initial and final state belong to the same excitonic series. This nomenclature reflects the state of the art in modeling this experiments, which is largely based on the hydrogen model for the exciton, which assumes rotational invariant Hamiltonian, and labels the excitonic states as 1s, 2s, 2p, i.e. in terms the principal quantum number and the angular momentum. The hydrogen model can be accurate for Wannier excitons, but it requires to be integrated with approaches which account for the underlying symmetries and/or topology of the band structure. Moreover, it may fail in systems with Frenkel like excitons, where the exciton wave-functions are strongly affected by the lattice symmetry\cite{galvani2016excitons}, and do not follow the standard hydrogen series. It is here important to underline that strongly bound excitons, which are stable at room temperature, are usually more Frenkel than Wannier like, but that often an exact distinction between these two kinds of excitons cannot be made. 

For these reasons, in this manuscript we propose a fully \ai\, approach to the study of pump and probe spectroscopy, which includes both inter- and intra-exciton transitions on an equal footing and takes fully into account lattice symmetries. We apply this approach to the case of a strong pump and a test probe in such a way as to compare the results with a perturbative approach based on quasi-equilibrium response theory. However, we expect that real-time propagation could be used beyond this regime also including pump-probe interference and dephasing effects which at present are not included in our approach.

Since two laser pulses are involved, TR-Abs can be described as a non-linear response of the material under the action of both the pump and the probe (P\&P) laser pulses, and an approach similar to the one developed to model two-photons absorption could be used.
The signal reconstructed by computing a non-linear response function is, by construction, perturbative in the effect of the pump pulse. 
      Furthermore, the approach can only be used to describe the situation where the probe detects the state directly generated by the pump, while it cannot deal with states generated by subsequent relaxation processes. In this situation, the physics of TR-Abs is very similar to that of equilibrium nonlinear response experiments. Extending nonlinear response theory beyond this regime becomes very cumbersome.\cite{virk2009multidimensional}
As an alternative, one can compute (or guess) the non-equilibrium state created the pump pulse and/or by the subsequent dynamics, and compute the non-equilibrium linear response with respect to the probe pulse.
In this second approach, the pump pulse can be considered in a non perturbative way, and deviations from the ideal excitonic picture can be computed. More importantly the approach holds beyond the overlap regime.
This angle clearly shows that TR-Abs experiments can access much more information, if compared to equilibrium non-linear optics experiments. Due to the action of relaxation processes, the initially generate bright excitonic state can be send into other dark excitonic states (also including finite momentum excitons). Thus both exciton to exciton transitions involving finite momentum excitons, and exciton dynamics can be probed in TR-abs setup.

In the present manuscript we follow the second approach outlined above, and employ a real-time propagation scheme based on an effective Schrodinger equation where correlation effects has been derived from Green's function theory~\cite{Attaccalite2011,Attaccalite2013}. In the equation of motion (EOM), the coupling term for the external field is constructed via the non-equilibrium Berry phase theory, where both the pump and the probe laser pulses are included.\cite{souza2004dynamics} This analysis does not rely on any excitonic basis set, and the excitonic state created by the pump pulse naturally emerges from the inclusion of the many-body self-energy. 
Moreover, in order to analyze the results obtained via the explicit real-time propagation scheme for the THz / infrared response function, we employ an alternative Fermi golden rule approach in the excitonic basis set. In this latter case, instead of computing the non-equilibrium state, we assume it can be described in terms of a well defined excitonic eigenstate obtained from the solution of the Bethe-Salpeter equation (BSE)~\cite{Strinati1988}. A similar approach has already been used in the study of the second harmonic response with very good results~\cite{riefer2017solving}. Moreover, the explicit use of excitonic basis-set allows us to analyze the symmetries of the excitonic wave-functions and use them to identify the dipole allowed exciton-exciton transitions, with an approach which generalizes the standard hydrogen model. While this approach holds in the low intensity limit, it also offers a starting point to describe deviation from the ideal excitonic picture, in terms of renormalization of the excitonic energies and wave-functions. 

We investigate two materials: bulk lithium fluoride (LiF) and monolayer hexagonal boron nitrite (hBN). LiF belongs to the family of alkali haliedes, and has been investigated in many theoretical and computational papers as a prototype material hosting tightly bound excitons~\cite{Shirley1996,Rohlfing1998,Puschnig2002,Wang2003,Marini2003,Abbamonte2008,Olovsson2009,ChiCheng2013,Gatti2013}.
hBN on the other hand is a layered insulator which in recent years has found many applications as a substrate and as a light emitter in the ultraviolet~\cite{schue2019bright}. From the theoretical point of view it has been extensively studied with both tight-binding models and \ai\ calculations~\cite{galvani2016excitons}, also including an accurate comparison with the hydrogen model labeling for the excitonic states.

The manuscript is organized as follows: in Sec.~\ref{sec:theory} we describe the theory for the real-time simulation, the Fermi golden rule approach and the selection rules that enter in the pump and probe spectroscopy; in Sec.~\ref{sec:results} we present P\&P response for LiF and hBN and discuss its interpretation; finally in Sec.~\ref{sec:conclusions} we draw conclusions and discuss perspectives from the theoretical and experimental point of view of P\&P techniques.

\section{Theory}
\label{sec:theory}
This section is dived in three parts. In Sec.~\ref{ssec:NEQ} we put forward the formalism to compute exciton-exciton transition. It is based on the non-equilibrium Berry Phase theory. The formalism holds at any exciton density, while the approximation we use for the self-energy, in particular the equilibrium screening approximation, limits the approach to a regime where the nonequilibrium screening can be neglected, i.e. below the Exciton-Mott transition. Within this limit, corrections to the spectra due to the increasing exciton density could be computed. Then in Sec.~\ref{ssec:FermiGolden} we discuss a simplified approach, based on the Fermi golden rule in the excitonic basis-set. This latter requires an initial guess for the many-body state generated by the pump pulse, and it holds only for low exciton densities. Starting from this, in Sec.~\ref{ssec:Diptrans}, we analyze the dipole allowed exciton to exciton transitions both in terms of the coefficients of the exciton wave-functions and taking advantage of group theory, starting from the point group associated to the crystal structure of the material.

\subsection{Real time propagation}\label{ssec:NEQ}
 In the real-time simulations the response of the system is obtained from the equations of motion for the valence band states:
 \begin{eqnarray}
	 i\hbar  \frac{d}{dt}| v_{m\bfk}(t) \rangle &=& \left[ H^{\text{MB}}_{\bfk} +i U^{ext}(t) \right] |v_{m\bfk}(t) \rangle \;, \label{eq:tdbse_shf}
\end{eqnarray}
where $| v_{m\textbf{k}}(t) \rangle$ is the periodic part of the occupied (at equilibrium) Bloch states. In the r.h.s. of Eq.~\eqref{eq:tdbse_shf}, $ H^{\text{MB}}_{\mathbf k}$ is the effective Hamiltonian derived from many-body theory~\cite{Attaccalite2011}.  
In order to catch excitonic effects in the real-time dynamics, we choose a many-body Hamiltonian $H^{MB}_{\mathbf k}$ in the form: 
\begin{equation}
  \label{eq:hmb}
  H^{MB}_{\mathbf k} \equiv H^{\text{KS}}_\bfk + \Delta H_\textbf{k} + \Delta\Sigma^{\text{HSEX}}[\rho(t)] \;.
\end{equation}
The unperturbed (zero-field) Hamiltonian is constructed starting from the \ks\ term, $H^{\text{KS}}_\textbf{k}$\cite{ks}, and includes the equilibrium quasi-particle corrections
$\Delta H_\textbf{k}$. The term $\Delta\Sigma^{\text{HSEX}}=\Sigma^{\text{HSEX}}[\rho(t)]-\Sigma^{\text{HSEX}}[\rho(0)]$ depends on the time dependent electronic density--matrix $\rho(t)$, which can be reconstructed, in the equilibrium basis set $| v_{m\textbf{k}}(0) \rangle$ from the time-dependent valence states as
\begin{equation}
	\label{eq:rho_eom}
\rho_{nm\bfk}(t) = \sum_{n'\in occ} \langle v_{n\bfk}(0) | v_{n'\bfk}(t) \rangle \langle v_{n'\bfk}(t) | v_{m\bfk}(0) \rangle ,
\end{equation}
where the sum runs over the occupied states only.
$\Delta\Sigma_{\text{HSEX}}$ describes the update, during the real-time propagation, of the Hartree (H) term, responsible for the local-field effects, plus of the Screened-EXchange (SEX) self-energy, that accounts for the electron-hole interaction and excitonic effect\cite{Strinati1988}.

The real-time propagation with the above Hamiltonian corresponds to a time-dependent (TD) version of the Bethe-Salpeter equation, and it is also known in the literature as TD-HSEX. Indeed, in the limit of small perturbation (\textit{i.e.} in the linear regime) these equations reproduce the optical absorption calculated with the standard $G_0W_0$ + BSE approach, as shown in Ref.~[\onlinecite{Attaccalite2011}].
Eq.~\eqref{eq:rho_eom} also shows how the EOM for the valence wave functions, e.g. eq.~\eqref{eq:tdbse_shf}, relates to the scheme using the EOM for the density matrix~\cite{Attaccalite2011}. In fact, while only valence block states are propagated in eq.~\eqref{eq:tdbse_shf}, $\rho_{nm\bfk}$ is expanded in a Kohn-Sham basis set, namely in terms of valence and conduction bands. Therefore, the number of conduction bands must be converged in both schemes~\cite{Sangalli2019}. However, knowing the time-dependent wave-function gives access to more information, and allows the calculation of the Berry polarization, as discussed below.

In the linear regime (and for non ferroelectric materials), a closed EOM can be written by expressing the coupling with the external field $U^{ext}(t)=\efield(t) \cdot \hat{\mathbf{P}}^{(1)}$, via the first order polarization operator $\hat{\mathbf{P}}^{(1)}$ written in the KS basis-set in terms of the transition dipoles $\bfr_{nm\bfk}$, e.g. $P^{(1)}_{nm\bfk}=-e\bfr_{nm\bfk}$. The expectation value of such operator gives the macroscopic polarization: 
\begin{equation}
\mathbf{P}^{(1)}=-e\tilde{\sum}_{n m\bfk}\bfr_{nm\bfk}\rho_{nm\bfk}
\label{eq:simpleP}
\end{equation}
where the $\sim$ on the sum means that all terms for which $|\epsilon_{n\bfk}-\epsilon_{m\bfk}|<\epsilon_{thresh}$ are neglected, since $\hat{P}^{(1)}_{nm\bfk}$ is not defined for those terms. Here $\epsilon_{n\bfk}$ are the eigenvalues of $H^{KS}_\bfk$ at equilibrium, and $\epsilon_{thresh}=10^{-5}$ eV (see also related discussion in sec.~\ref{ssec:FermiGolden}).
However, in the present case we need to probe the response to the system after the action of the pump and we thus need to go beyond this first order approach. This can be achieved by means of the Berry Phase formulation, e.g.
by defining the coupling with the external field as $U^{ext}(t)=\efield(t) \cdot \tilde \partial_\textbf{k}$, which holds to all orders. As we imposed Born-von K\'arm\'an periodic boundary conditions, the coupling takes the form of a $\textbf{k}$-derivative operator $\tilde \partial_\textbf{k}$. The tilde indicates that the operator is ``gauge covariant'' and guarantees that the solutions of Eq.~\eqref{eq:tdbse_shf} are invariant under unitary rotations among occupied states at $\textbf{k}$ (see Ref.~[\onlinecite{souza2004dynamics}] for more details). This derivative is calculated using a finite-difference five-point midpoint formula \cite{nunes2001berry}.  


From the evolution of $| v_{m\textbf{k}} \rangle$ in Eq.~\eqref{eq:tdbse_shf} we calculate the real-time polarization $\text{P}_{\mathbf{a}i}$  along the lattice vector $\mathbf{ a}_i$ as
 \begin{equation}
	 \text{P}_{\mathbf{a}i} = -\frac{ef |\mathbf{a}_i| }{2 \pi \Omega_c} \text{Im log} \prod_{\textbf{k}}^{N_{\textbf{k}}-1}\ \text{det} S\left(\textbf{k} , \textbf{k} + \mathbf q_i\right)\label{eq:berryP}  \;, 
 \end{equation}
where $S(\textbf{k} , \textbf{k} + \mathbf q_i) $ is the overlap matrix between the valence states $|v_{n\textbf{k}}\rangle$ and $|v_{m\textbf{k} + \textbf{q}_i}\rangle$, $\Omega_c$ is the unit cell volume,  $f$ is the spin degeneracy, $N_{\textbf{k}}$ is the number of $\textbf{k}$ points along the polarization direction, and $|\mathbf{q}_i| = 2\pi/(N_{\textbf{k}} |{\mathbf{a}_i}|)$.
Eq.~\eqref{eq:berryP} is the adiabatic extension to the non equilibrium regime of the definition of the polarization in an extended system~\cite{souza2004dynamics}.

Via the solution of Eq.~\eqref{eq:tdbse_shf}, the transient absorption signal is finally defined as the Fourier transform of
\begin{equation}
\Delta \mathbf{P}(t)=\mathbf{P}_{pp}(t)-\mathbf{P}_{p}(t)
\end{equation}
where $\mathbf{P}_{p}(t)$ is the real-time polarization generated by the action of the pump pulse alone, while $\mathbf{P}_{pp}(t)$  is the real-time polarization generated by the action of both the pump and the probe laser pulse.
No dephasing mechanism is included in the simulation, in order to preserve the coherent excitonic state generated by the pump laser pulse. While decoherence is expected to happen in experiments, it is far from trivial to formulate a proper decoherence term in the Eq.~\eqref{eq:rho_eom}. On the contrary it was shown that by dephasing the one-body density matrix, i.e. sending to zero the off-diagonal elements, leads to a non coherent state which involves non-bound electron-hole pairs, thus destroying not only coherence, but also the excitonic state.~\cite{Perfetto2019b,Sangalli2021}
Instead, a finite broadening parameter $\eta$ is added later, when performing the Fourier transform of the polarization
\begin{equation}
	\label{eq:P_fft}
	\Delta \bfP(\omega,t_p)=\int_{t_{p}}^{+\infty} dt\, \Delta \bfP(t) e^{i\omega t - \eta (t-t_{p})}.
\end{equation}
where $t_{p}$ is the starting time of the probe field. Finally the non-equilibrium response function is obtained~\cite{Otobe2016,Perfetto2015} dividing by the probe pulse: ${\chi_{\boldsymbol{\mu}\boldsymbol{\mu}}(\omega,t_p)=\Delta \bfP(\omega,t_p)/\bfE_p(\omega)}$ .
The $\eta$ parameter defines the broadening of the peaks in the response function.
The approach can be used to model transient-absorption experiments 
both in the low energy range (typically THz of infrared) where exciton to exciton transitions are expected, and in the resonant energy range, where shifts and changes to the equilibrium absorption peaks are expected. In the present manuscript we focus on the low energy range. Since we have the static HSEX self-energy and we do not include any additional decoherence (or scattering term), the system will remain in the state created by the pump pulse and the spectra will be independent from $t_p$. One of the scopes of the present manuscript is to show that exciton-exciton transition can be captured within TD-HSEX.

\subsection{Non equilibrium response function} \label{ssec:FermiGolden}
In order to analyze the results from the real time propagation scheme it is useful to formulate a simplified approach in terms of the excitonic basis set.
We start from the general expression for a ``A operator''-``B operator'' linear response function~\cite{Chuang1991}
\begin{equation}
\label{eq:Chi_def1}
\x_{AB}(\w) = \frac{2}{V}\sum_{J\neq I}
\frac{A_{IJ} B_{JI}}{(E_J-E_I)-\w-i\eta} \, .
\end{equation}
Here the $I,J$ indexes represent the initial, $\ket{I}$, and final, $\ket{J}$,  many--body states states with energies $E_I$ and $E_J$, and $A_{IJ}$ and $B_{IJ}$ the matrix elements of some operator. The dipole-dipole response function is obtained for ${A=\mu^{\alpha}}$, ${B=\mu^{\beta}}$, with the $\alpha$-th and $\beta$-th cartesian component of the matrix element of the many body dipole operator $\boldsymbol{\mu}_{IJ} = \bra{I}\dipole\ket{J}$. 
The current current response function instead is obtained for ${A=j^{\alpha}}$, ${B=j^{\beta}}$, with the $\alpha$-th and $\beta$-th cartesian component of the matrix element of the many body velocity operator $\bfj_{IJ} = \bra{I}\hat{j}\ket{J}$. Both $\chi_{\mu_\alpha\mu_\beta}$ and $\chi_{j_\alpha j_\beta}$ can be used to define the optical absorption of a material.  $\epsilon(\omega)=1-4\pi\chi_{\mu_{\alpha}\mu_{\beta}}(\omega)$ in the length gauge, or
$\epsilon(\omega)=1-4\pi\chi_{j_{\alpha}j_{\beta}}(\omega)/\omega^2$ in the velocity gauge. In the latter case we explicitly remove the divergent $1/\omega^2$ and $1/\omega$ terms by expanding $\chi_{j_{\alpha}j_{\beta}}(\omega)/\omega^2$ and imposing sum rules~\cite{sangalli2017optical}. 

At equilibrium the initial state $\ket{I}=\ket{g}$ is the ground state of the system with $E_I=E_0$, and the final state is an exciton state $\ket{J}=\ket{\l\bfq}$ with $E_J=E_0+\omega_\lambda(\bfq)$.
Here $\bfq$ is the excitonic momentum index. Assuming that the ground state has zero momentum (i.e. no charge density wave or spin density wave ground state), only zero momentum excitons needs to be considered.
In this case the expression of the response function reduces to 
\begin{eqnarray}
\label{eq:Chi_def_0}
\x^{eq}_{AB}(\w)
&=& \frac{2}{V}\sum_{\l}\frac{A_{0\l}(\mathbf{0}) B_{\l0}(\mathbf{0})}{\omega_\l(\mathbf{0})-\w-i\eta} \, .
\end{eqnarray}
Within the Tamm-Dancoff approximation (TDA), the exciton states can be expressed as a linear combination of valence-conduction pairs, that is
\begin{equation}
\label{eq:exciton_def}
\ket{\l\bfq} = \sum_{cv\bfk} A_{cv\bfk}^{\l\bfq}
\ket{c\bfk-\bfq}\otimes\ket{v\bfk}  \, .
\end{equation}
Here the $c$ and $v$ indexes run over conduction and valence states, respectively, while $\bfk$ is the electronic momentum index. The response function obtained by inserting Eq.~\eqref{eq:exciton_def} into Eq.~\eqref{eq:Chi_def_0} is identical to the one obtained via a formal solution of the many-body response function via many-body perturbation theory (MBPT) within the TDA.
It can be shown that the dipole matrix elements $\boldsymbol{\mu}_{0\lambda}(\bfq)=\bra{\l\bfq}\dipole\ket{g}$ reduce to a linear combination of single-particle terms (see appendix A)
\begin{equation}
\label{eq:dipole_1part}
\boldsymbol{\mu}_{0\lambda}(\bfq)= e\, \d(\bfq) \sum_{cv\bfk} A_{cv\bfk}^{\l\mathbf{0}} \bra{c\bfk}\opr\ket{v\bfk} \, .
\end{equation}
with $e$ the electronic charge and $\opr$ the one body position operator.
The assumption that the ground state has zero momentum reflects in the $\d(\bfq)$ function~\footnote{Here $\bfq$ is the momentum of the final state, while we are interested in optical transitions only, i.e. zero transferred momentum $\bfQ$. Different operators can be used to define transitions with finite momentum transfer in the length and in the velocity gauge, namely $e^{i\bfQ\cdot\bfr}$ and $\hat{v} e^{i\bfQ\cdot\bfr}$. We do not discuss these transitions here.}. The corresponding expression for the matrix elements of the velocity operator $\bfj_{0\lambda}(\bfq)=\bra{\l\bfq}\hat{\bfj}\ket{g}$ can be 
defined by dividing and multiplying by the transition energies~\cite{sangalli2017optical}. Neglecting this
ratio provides an approximated expression: 
\begin{eqnarray}
\label{eq:dipole_1part_vel}
\bfj_{0\lambda}(\bfq)
&=& e\, \d(\bfq) \sum_{cv\bfk} A_{cv\bfk}^{\l\mathbf{0}} \frac{\bra{c\bfk}\opv\ket{v\bfk}}{\Delta\epsilon_{cv\bfk}}E_\lambda(\mathbf{0}) \, ,
\\
&\approx& e\, \d(\bfq) \sum_{cv\bfk} A_{cv\bfk}^{\l\mathbf{0}} \bra{c\bfk}\opv\ket{v\bfk} \, . 
\end{eqnarray}
where $\Delta\epsilon_{cv\bfk}=\epsilon_{c\bfk}-\epsilon_{v\bfk}$.
The error induced amounts to a re-normalization of the peaks intensity~\cite{sangalli2017optical} without significant changes in the equilibrium absorption. 
Eq.~\eqref{eq:dipole_1part_vel} cannot be used when $\Delta\epsilon_{nm\bfk}<\epsilon_{thresh}$ since the ratio $\bra{c\bfk}\opv\ket{v\bfk}/\Delta\epsilon_{cv\bfk}$ becomes numerically unstable. This is also related to the fact that intra-band matrix elements $\bra{n\bfk}\opr\ket{n\bfk}=\bra{n\bfk}\opv\ket{n\bfk}/\Delta\epsilon_{nn\bfk}$ are ill defined since $\Delta\epsilon_{nn\bfk}=0$. In our code $\epsilon_{thresh}=10^{-5}$ eV. While at equilibrium this never happens in systems with a gap, transitions with $\Delta\epsilon_{nm\bfk}<\epsilon_{thresh}$ will be involved in the TR-abs spectrum (see later discussion around Eq.~\eqref{eq:velocity_dipole_exc_complete}).

In the non-equilibrium regime, we assume that an excitonic state has been created by the pump laser pulse, $\ket{I}=\ket{\lambda_i\bfq}$ with $E_I=E_0+\omega_{\lambda_i}(\bfq)$, where the superscript ``$i$" is used to highlight that this is the initial excitonic state.
The non-equilibrium dipole-dipole response function in the energy range of inter-exciton transitions takes the form (see also App.~\ref{App:DipSecRules})
\begin{eqnarray}
\label{eq:Chi_def_exc}
\x^{\lambda_i\bfq}_{\mu_\alpha\mu_\beta}(\w)
&=& \frac{2}{V} \sum_{\l}\frac{\mu_{\l_i\l}^\alpha(\bfq) \mu_{\l\l_i}^\beta(\bfq)}{\omega_\l(\bfq)-\omega_{\l_i}(\bfq)-\w-i\eta}
\end{eqnarray}
Eq.~\ref{eq:Chi_def_exc} holds at low excitonic densities because we are assuming that the equilibrium excitonic energies and wave-functions can be used, and because no final state population effects are considered 
\footnote{While the formulation in terms of real time propagation can account for these effects, a full formulation in terms of the response function could be also achieved within equilibrium MBPT formalism. This would require to apply MBPT starting from a non interacting excited state\cite{Perfetto2019b}. This however leads to problems, since the formalism would rely on the use of the adiabatic connection for excited states.}.
The intensity of the transient absorption signal is weighted by the exciton density injected by the pump pulse. 

We observe that for optically injected excitons the condition $\bfq=\mathbf{0}$ holds, however
dissipation and relaxation mechanisms could scatter the initial state into finite momentum excitonic states leading to a population $N_{\lambda_i}({\bfq})$, and the overall transition signal can be expressed as
\begin{equation}
\label{eq:Chi_def_neq}
\x_{\mu_\alpha\mu_\beta}(\w)=\sum_{\lambda_i\bfq} N_{\lambda_i}(\bfq)\x^{\lambda_i\bfq}_{\mu_\alpha\mu_\beta}(\w)
\end{equation}
This is why in general, even in presence of an optical pump and an optical probe, finite momentum excitons need to be considered to model inter-excitons transitions. The above expression assumes that the non coherent populations do no give rise to interference patterns in the transient absorption signal.
In the present manuscript, we do not account for relaxation and dissipation mechanisms and we will focus on the situation where $N_{\lambda_i}(\bfq)=N^{exc}\d_{\l_i,\l_0}\d(\bfq)$.

\subsection{Selection rules}
\label{ssec:Diptrans}
The crucial step is then to identify the dipole allowed $\lambda \rightarrow \lambda'$ transitions. 
Starting from the hydrogen model, and dividing exciton-exciton transitions into intra- and inter-exciton transition~\cite{Jorger2005}, selection rules for the intra-exciton transitions can then be established starting from the quantum numbers of the hydrogen model used to describe the exciton. 
However, this approach neglects the symmetries of the crystal in the excitonic envelop, and it is a good approximation only for delocalized Wannier excitons. For more localized Frenkel excitons, a fully \ai\ approach, and/or an approach which accounts for the symmetries of the lattice, offers a better description.

\subsubsection{Ab-initio dipole matrix elements}
\label{ssec:DiptransAi}
In the \ai\ formalism, selection rules can be computed explicitly by defining the expression for the dipole matrix elements $\boldsymbol{\mu}_{\lambda\lambda_i}(\bfq)=\bra{\l\bfq}\dipole\ket{\l_i\bfq}$ (see derivation in App.~\ref{App:ExcExcDips}):
\begin{widetext}
\begin{equation}
	\label{eq:dipole_exc_complete}
 \boldsymbol{\mu}_{\lambda_i\lambda}(\bfq) = 
\sum_{v,cc',\bfk}\(A^{\l\bfq}_{cv\bfk}\)^*A^{\l_i\bfq}_{c'v\bfk}\bfr_{cc'\bfk} 
-\sum_{c, vv',\bfk}\(A^{\l\bfq}_{cv\bfk}\)^*A^{\l_i\bfq}_{cv'\bfk}\bfr_{v'v\bfk-\bfq}  
+\sum_{v,c,\bfk}\(A^{\l\bfq}_{cv\bfk}\)^*A^{\l_i\bfq}_{cv\bfk} \sum_{n\in occ}\bfr_{nn\bfk}
\, .
\end{equation}
\end{widetext}
The dipole matrix elements depend both on the excitonic coefficients $A^{\l\bfq}_{cv\bfk}$  and on the 
interband dipoles $r_{nm\bfk}$ over all the $\bfk$-points of the BZ. 
Eq.~\eqref{eq:dipole_exc_complete} is a generalization of the simplified analysis of~Ref.~\onlinecite{Poellmann2015}, 
that holds for delocalizaed Wannier excitons, in which the dipole matrix element $\boldsymbol{\mu}_{\lambda_i\lambda}(\bfq)$ depends only on the excitonic wave-functions and their quantum numbers in the hydrogenic model. 

The first two addends of \eqref{eq:dipole_exc_complete} have a very simple interpretation in terms of transitions from state $\lambda_i$ to state $\lambda$ mediated by the electronic dipole. Their expression is very similar to the one of the exciton-phonon matrix elements~\cite{antonius2022theory}, although it involves only one momentum, since the initial and final exciton must have the same momentum.
However an issue appears since Eq.~\eqref{eq:dipole_exc_complete} involves the intra-band dipoles $\mathbf{r}_{nn\bfk}$ which are ill defined in periodic boundary conditions. 
This points to the fact that we are looking to a non-linear response, as discussed in the introduction. 
Moving to the velocity gauge and to the current-current response function solves this issue, since intra-band velocity dipoles are well defined
\begin{multline}
\label{eq:velocity_dipole_exc_complete}
\boldsymbol{\bfj}_{\lambda_i\lambda}(\bfq) \approx 
\sum_{v,cc',\bfk}\(A^{\l\bfq}_{cv\bfk}\)^*A^{\l_i\bfq}_{c'v\bfk}\bfv_{cc'\bfk-\bfq} \\
-\sum_{c, vv',\bfk}\(A^{\l\bfq}_{cv\bfk}\)^*A^{\l_i\bfq}_{cv'\bfk}\bfv_{v'v\bfk}  \\
+\sum_{v,c,\bfk}\(A^{\l\bfq}_{cv\bfk}\)^*A^{\l_i\bfq}_{cv\bfk} \sum_{n\in occ}\bfv_{nn\bfk}
\, . 
\end{multline} 
The velocity gauge introduces an overall error in the intensity of the dipoles (similarly to the equilibrium case, see previous discussion related to Eq.~\eqref{eq:dipole_1part_vel})~\cite{sangalli2017optical}. 
At variance with the equilibrium case, however, we cannot divide by the transition energies $\Delta\epsilon_{nn\bfk}$ and later multiply by the excitonic energies, since this operation  is numerical unstable for $\Delta\epsilon_{nn\bfk}<\epsilon_{thresh}$ and ill defined for the case $\Delta\epsilon_{nn\bfk}=0$. Thus, we use Eq.~\eqref{eq:velocity_dipole_exc_complete} when computing the spectra in the velocity gauge, which in short, introduces an overall error in the intensities, but accounts for the role of intra-band transitions.

Finally, we underline that in presence of degenerate excitonic states it is crucial to identify a specific state in the degenerate subspace to describe the exciton generated by the pump pulse. The polarization of the selected exciton must be parallel to the field polarization projected into the polarization space spanned by the degenerate excitons. This allows us to capture the dependence of the transient absorption spectrum on the relative orientation of the pump and the probe pulse polarization. In App.~\ref{App:DegenerateExc} we show how this dependence can be taken into account.

\subsubsection{Symmetry considerations} \label{ssec:DiptransSymm}

In support to the real-time calculations and the non-equilibrium response functions, we provide a symmetry analysis of the different spectra presented in this work. In particular we analyze the matrix elements
\begin{equation}
\label{eq:DipoleExcSymm}
\boldsymbol{\mu}_{IF} = \bra{F}\dipole\ket{I} 
\end{equation}
that enter in Eq.~\ref{eq:Chi_def_exc} using group theory\cite{Dresselhaus2008} in such way to determine which matrix elements are zero by symmetry and which are not.
To do this, we assign each of the three elements in Eq.~\eqref{eq:DipoleExcSymm}, $\bra{F}$, $\dipole$, and $\ket{I}$, the irreducible representation(irreps) to which it belongs, with respect to the symmetry group of the crystal. Let us call them $O_F$, $O_\mu$, $O_I$. Then the integral is different from zero \textit{if and only if} the identity element belongs to the product of the three symmetry operations~\cite{Dresselhaus2008}. This result can be achieved using the product table of the symmetry point group and furthermore
also polarization effects can be taken into account by considering the ``partner''  (here we use the same nomenclature of Ref.~\onlinecite{Dresselhaus2008}) of each irrep associated to a specific polarization. This analysis is performed in details for the point groups of the two materials considered in App.~\ref{App:DipSecRules} of the present manuscript.

\section{Results}
\label{sec:results}

In this section we will compute the Tr-Abs properties of LiF and hBN usin the two approaches discussed in the previous section. We will label ``TD-HSEX'' results the one obtained from the real-time propagation scheme, and ``Exc. Fermi'' results based on the Fermi Golden rule in the excitonic space. In particular ``Exc. Fermi (length)'' if the TR-Abs is constructed via eq.~\eqref{eq:dipole_exc_complete} and ``Exc. Fermi (velocity)'' via eq.~\eqref{eq:velocity_dipole_exc_complete}. It is worth to mention that, while the TD-HSEX scheme describes exciton-exciton transition starting from a coherent excitonic state, the Exc. Fermi approach describes exciton-exciton transition starting from a non-coherent excitonic population. However, as already shown for the case of ARPES, the two give rise to the same signal, provided that the average population of the coherent state is equal to the population of the non coherent state. Eventual differences due to the coherent polarization associated with the excitonic state, such as Franz-Keldysh oscillations in the TR-Abs spectra, could be observed only in specific conditions~\cite{Otobe2016}. This is beyond the scope of the present manuscript.

\subsection{Results on LiF}
\begin{figure}[t]
\begin{center}
\includegraphics[width=0.45\textwidth]{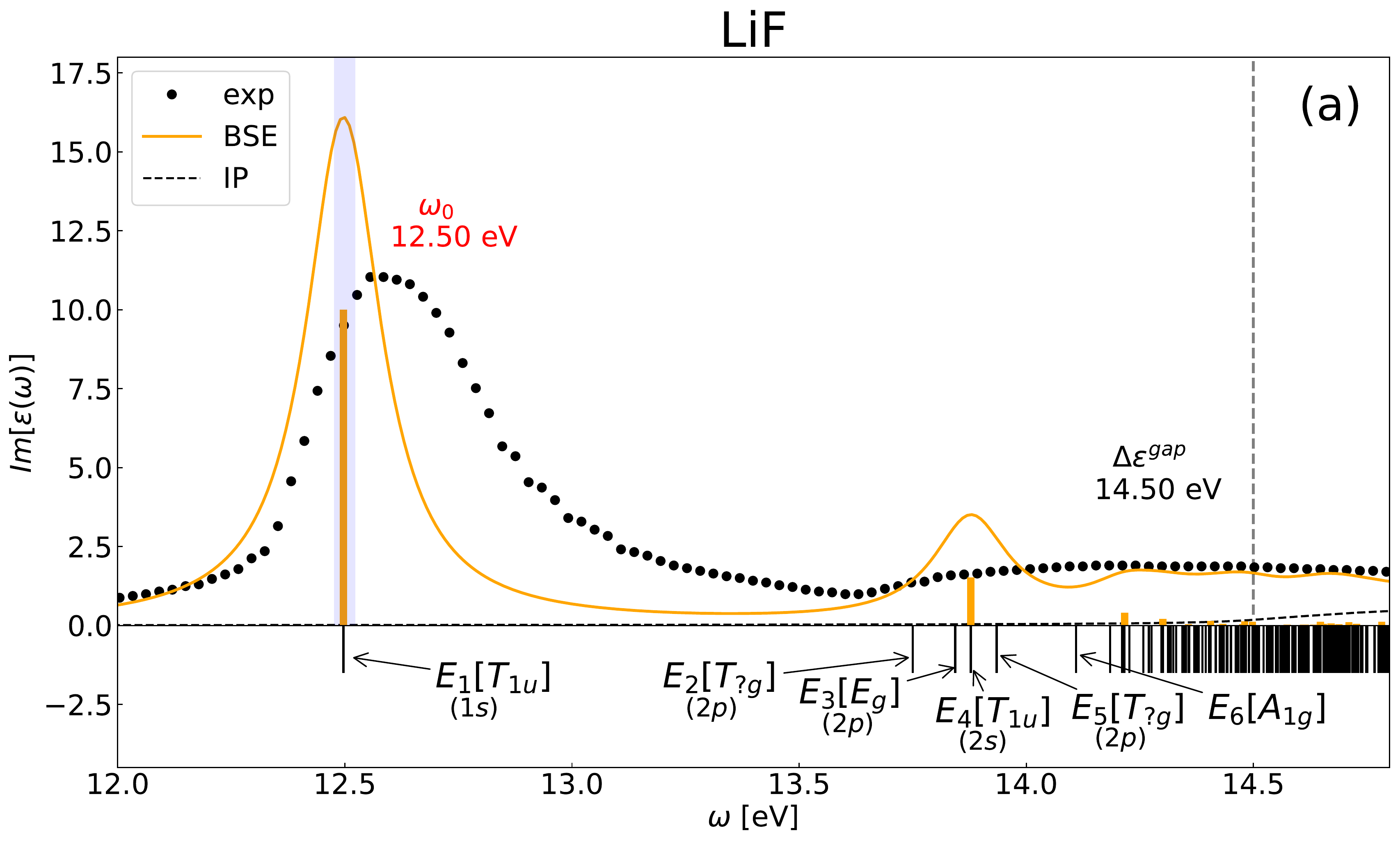}
\includegraphics[width=0.45\textwidth]{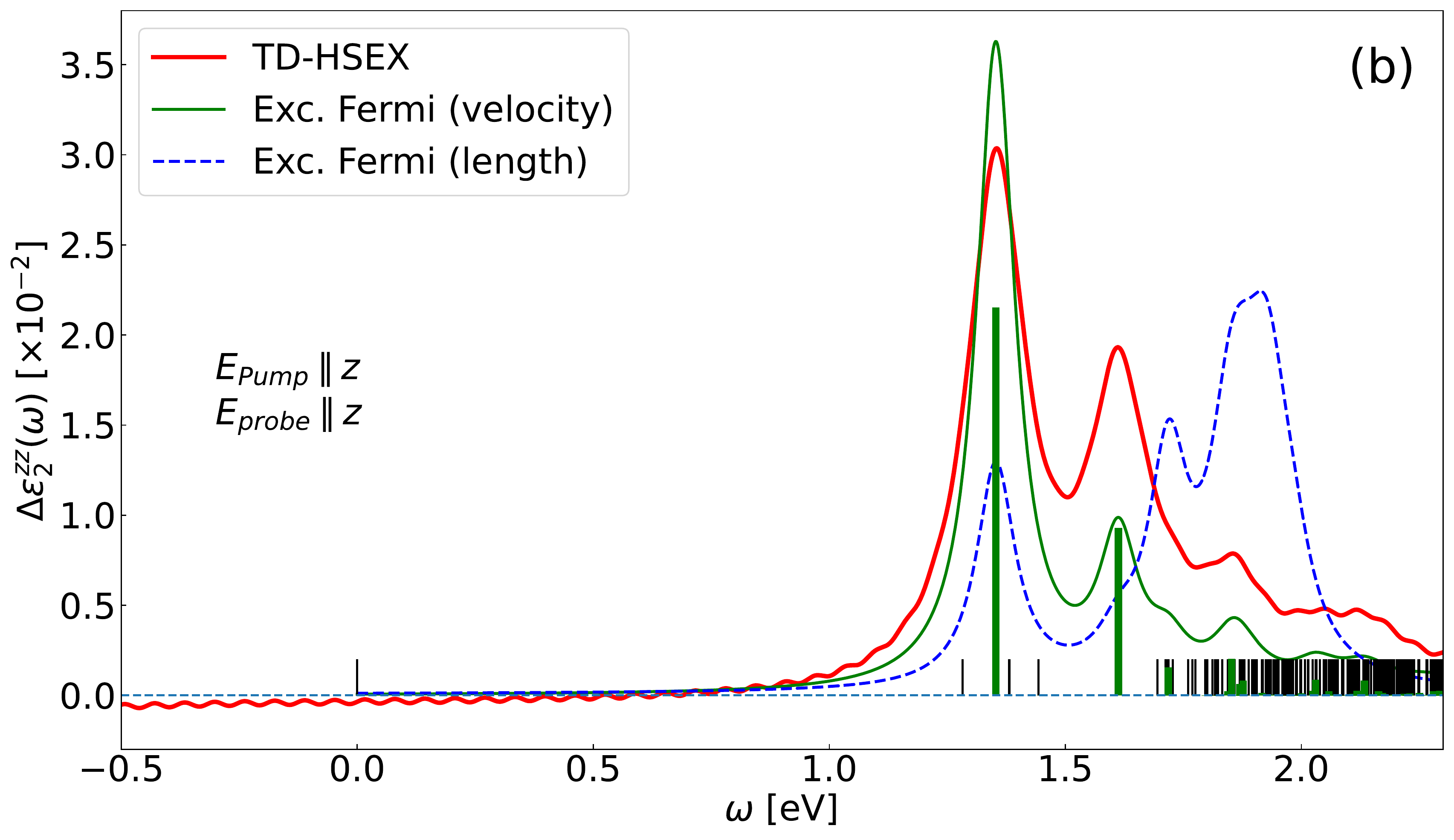}
\includegraphics[width=0.45\textwidth]{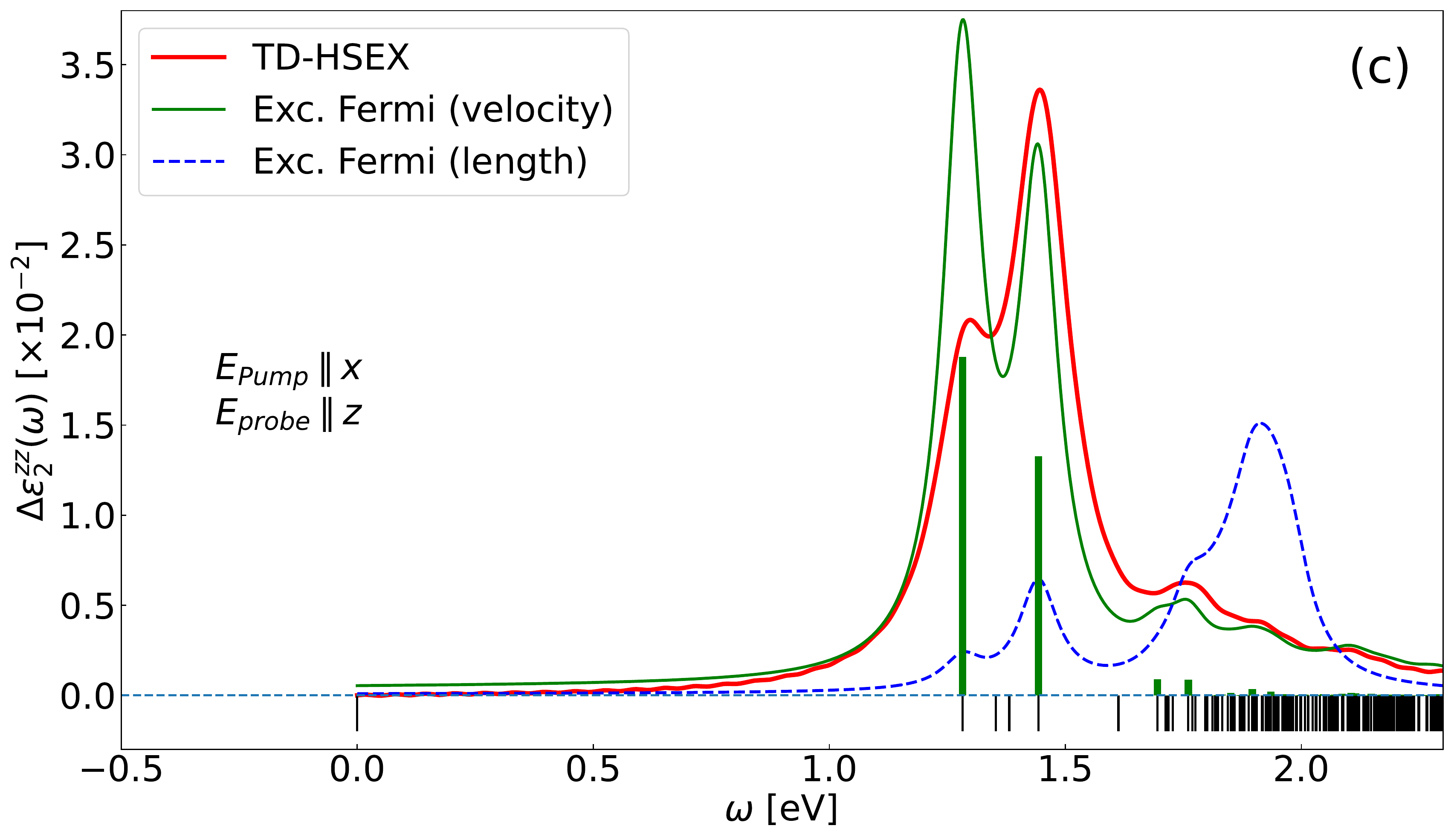}
\end{center}
\caption{(a) Absorption of LiF. All poles $\omega_\lambda$ are represented as vertical black lines. Orange vertical bars are proportional to $|\boldsymbol{\mu}_{0\lambda}(\mathbf{0})|^2$ (b-c) Transient Absorption of LiF after pumping the system with a pump resonant with the first excitonic pole $\omega_1$. All possible transitions $\Delta\omega=\omega_i-\omega_1$ are represented with vertical black lines. Green vertical bars are proportional to $|\bfj_{\lambda\lambda_i}(\mathbf{0})|^2/\omega^2$. (b) Configuration with probe parallel to the pump. (c) Configuration with probe perpendicular to the pump.}
\label{fig:abs1}
\end{figure}

LiF is a wide gap insulator, with an electronic band gap of $\approx 14.5$ eV (experimental data give a gap in the range 14.1-14.5 eV, while GW simulation on top of LDA a value close to 14.4 eV~\cite{Shirley1996}). The absorption spectrum is dominated by an intense excitonic peak ($E_1$) at $\approx 12.5$ eV (see the panel a of Fig.~\ref{fig:abs1}). $E_1$ has been often described as a charge transfer (CT) exciton, involving transfer of an electron from the alkaline atom (Li) $p$ to a halogen atom (F) $s$ level~\cite{Abbamonte2008}. This description emerges from the fact that LiF is a ionic crystal, in which the lone Li($2s$) electron of the isolated Li atom is transferred to the empty F($2p$) level of the F atom. Accordingly, the valence band structure is mainly composed by the F($2p$) orbitals, while the lowest conduction band is the Li($2s$) states. 
$E_1$ has been studied in different works and, while initially proposed as intermediate between Frenkel and Wannier, it has more recently identified as a strongly bound Frenkel exciton~\cite{Abbamonte2008}.
Due to its large binding energy of $\approx 2$ eV, it has often been used to validate theoretical development. For example its wave-function~\cite{Rohlfing1998,Rohlfing2000} and dispersion~\cite{Gatti2013} were analyzed in details. Other excitons, involving a semi-core hole, in the XUV energy range, have been investigated in the literature by means of \ai\, simulations~\cite{Olovsson2009}, reporting even larger binding energies.

Instead, less attention has been payed to the excitons, either dark or bright, which lie in the energy window between $E_1$ and the electronic gap, and which characterize the excitonic series of LiF. In the present work we investigate these excitons which, after the action of a laser pulse tuned resonant with $E_1$, dominate the transient absorption spectrum of LiF in the energy range from 0 to 2 eV.
We label them $E_2, E_3, E_4, ... $.

We employ the standard GW+BSE scheme on top of LDA. Ground state LDA calculation are performed with norm conserving pseudo-potentials, an energy cutoff of 80 Ry for the wave-functions and a 6x6x6 k-points grid. RPA screening is computed with cutoff of 8 Ry for the response function $\chi_{\bfG\bfG'}(\bfq,\omega)$ and using 50 bands in the internal sum over the conduction states. Finally the BSE calculation with a scissor or 6.05 eV, bands from 2 to 5, k-points grid 24x24x24, energy cutoff 8 Ry for the screened e-h interaction $W_{\bfG\bfG'}$, and 32 Ry for the eh exchange interaction $v_{\bfG}$. The same parameters are then used for the real-time propagation within the TD-HSEX simulations, with the bands 2 to 5 entering the bands indexes in Eq.~\eqref{eq:rho_eom} and in Eqs.~\eqref{eq:H_kernel}-\eqref{eq:SEX_kernel}, k-points grid 24x24x24, 8 Ry for $W_{\bfG\bfG'}$, and 32 Ry for $v_{\bfG}$. Moreover a time step of $10^{-2}$ femto-seconds (fs) is used to time integrate Eq.~\eqref{eq:rho_eom} with the Crank-Nicolson integrator for a total time of 100 fs, which is later used to integrate Eq.~\eqref{eq:P_fft}.

In Fig.~\ref{fig:abs1} absorption and transient absorption of LiF are shown. The solution of the excitonic matrix shows several bound poles, many of which are dark. The first 6 poles are well separated from the others. Besides the well characterized lowest energy bright exciton $E_1$, also $E_4$ is bright, while the other 4 are dark, see panel a of Fig.~\ref{fig:abs1}. The position of the bright exciton $E_4$, around 13.9 eV, seems to be confirmed by the shape of the experimental absorption which bounces back at around 13.7 eV, although with a very broad signal. All bright poles are three fold-degenerate, due to the symmetries of the crystal, while dark poles can have different degeneracy. The poles from $E_2$ to $E_5$ are located in the energy range between 13.78 and 13.94 eV. $E_2$ and $E_5$ are also three-fold degenerate, while $E_3$ is 
two-fold degenerate. Finally $E_6$ is shifted from the previous group, at 14.1 eV and it is non degenerate.

Turning our attention from equilibrium absorption to transient absorption  in the inter-exciton transitions regime, first of all we observe that the TD-HSEX scheme gives poles exactly at the energy differences $E_i-E_1$, with $i$ running over the other excitons. Such result can be obtained only using the Berry phase expression for the polarization, eq.~\eqref{eq:berryP}. Using instead TD-HSEX with eq.~\eqref{eq:simpleP} gives a zero spectrum in this energy range, regardless of the $\epsilon_{thresh}$ value. Moreover, it is remarkable that, after pumping the system in resonance with the lowest energy excitons, the 4 dark poles above mentioned can be detected by considering two different configurations: probe parallel to the pump pulse (Fig.~\ref{fig:abs1} panel b) and probe perpendicular to the pump pulse (Fig.~\ref{fig:abs1} panel c). This result shows that transient absorption is a very powerful technique to explore dark excitonic poles.
In both the configurations we show the transient absorption spectra computed by full real time propagation scheme at low pump intensity (red continuous line) and the spectra obtained within the ``excitonic Fermi golden rule'', either in the length gauge (blue dashed line) or in the velocity gauge (green continuous line). As discussed in the theory section, both gauges have some issues and are not able to precisely reproduce the results of the full real-time propagation. Indeed both blue and green lines correctly foresee which inter-exciton transitions are expected to be bright, but somehow they miss the correct relative intensity. The problem is much more severe in the length gauge which strongly underestimate the intensity of the low energy peaks and overestimates the intensity of the higher energy ones. The velocity gauge tends to do the opposite, but the agreement with the full real time propagation is much better and, overall, the velocity gauge approach can be used to obtain a good description of the spectra.

\begin{figure*}[t]
\begin{center}
\includegraphics[width=0.19\textwidth]{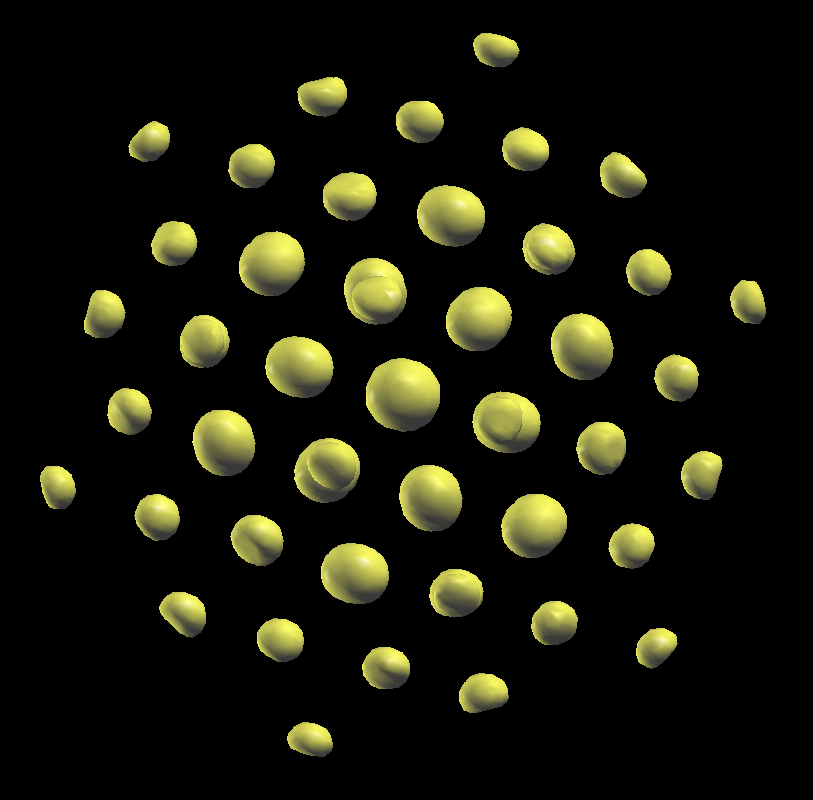}
\includegraphics[width=0.1875\textwidth]{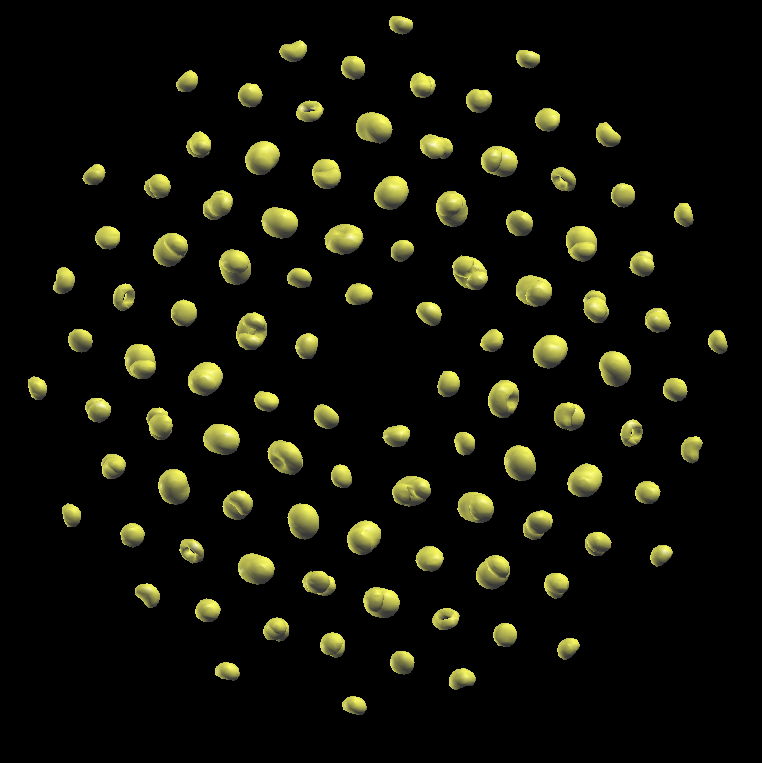}
\includegraphics[width=0.205\textwidth]{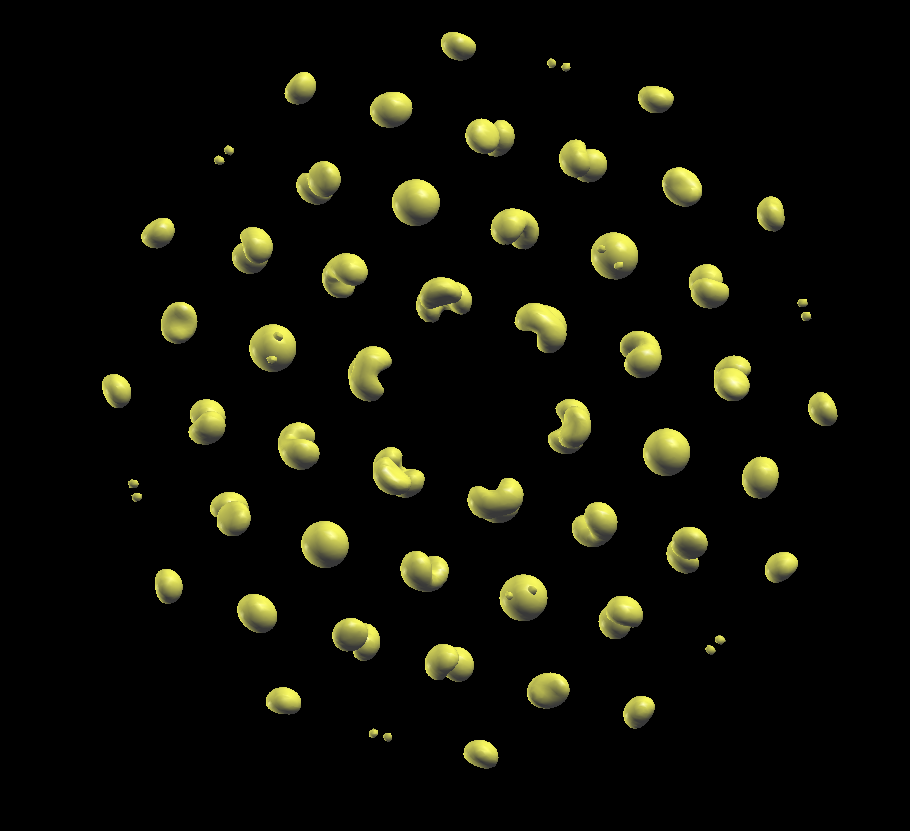}
\includegraphics[width=0.195\textwidth]{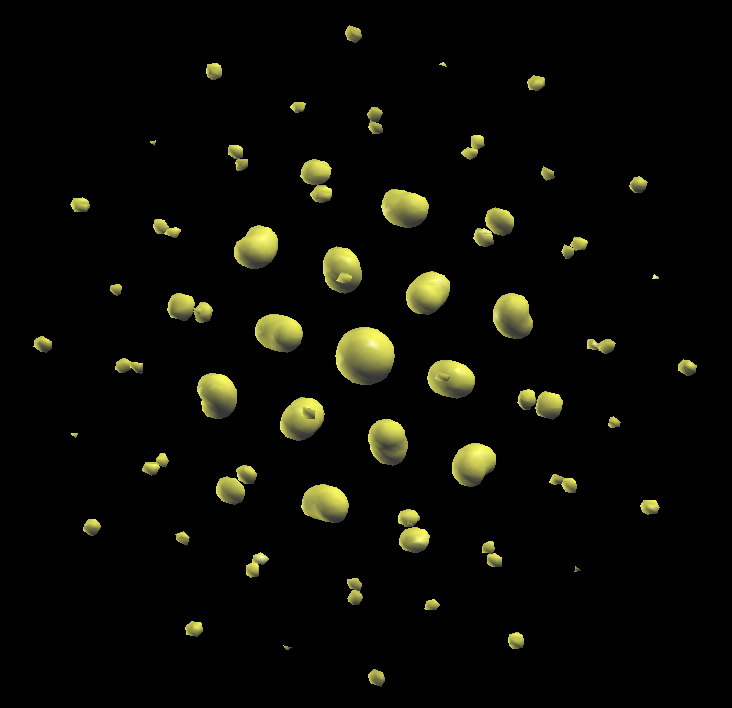}
\includegraphics[width=0.1875\textwidth]{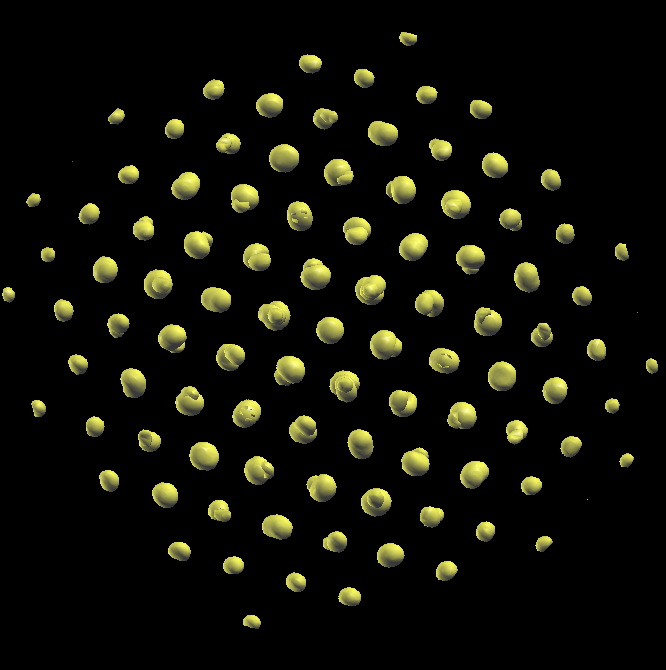}
\includegraphics[width=0.193\textwidth]{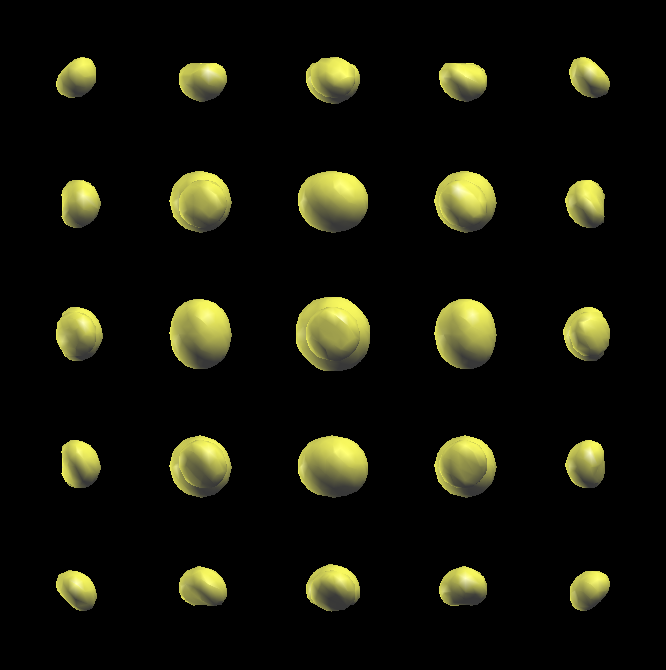}
\includegraphics[width=0.193\textwidth]{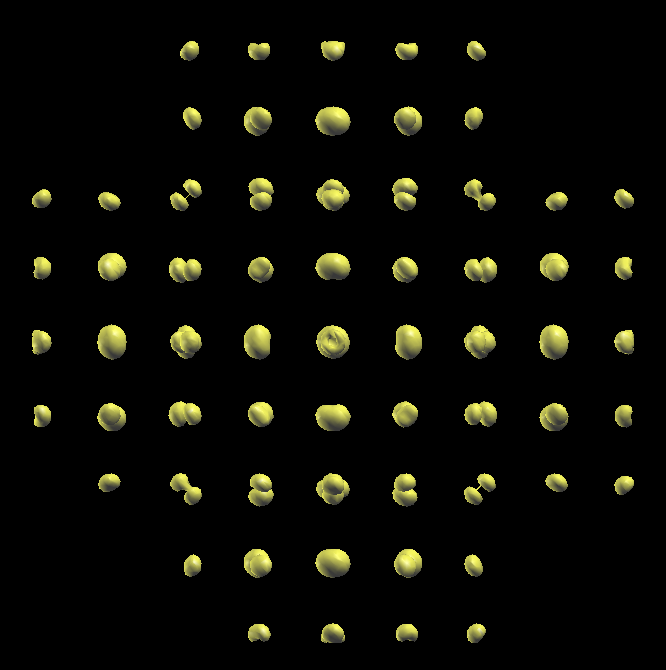}
\includegraphics[width=0.193\textwidth]{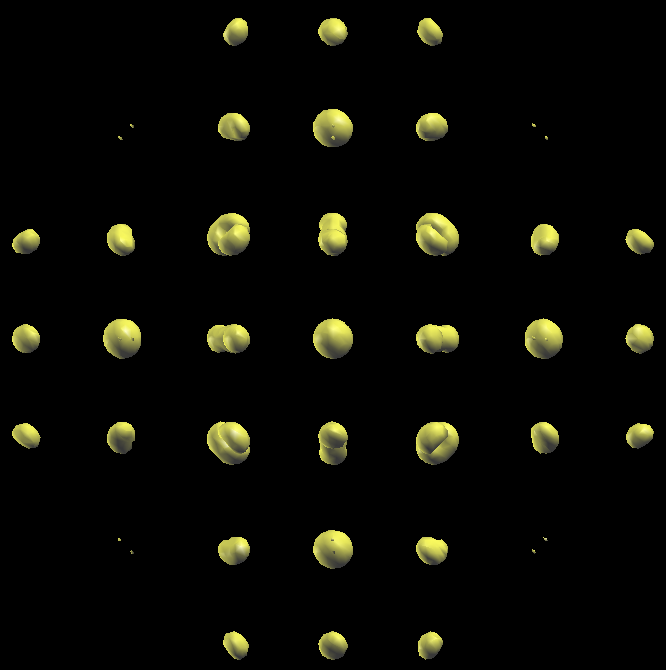}
\includegraphics[width=0.193\textwidth]{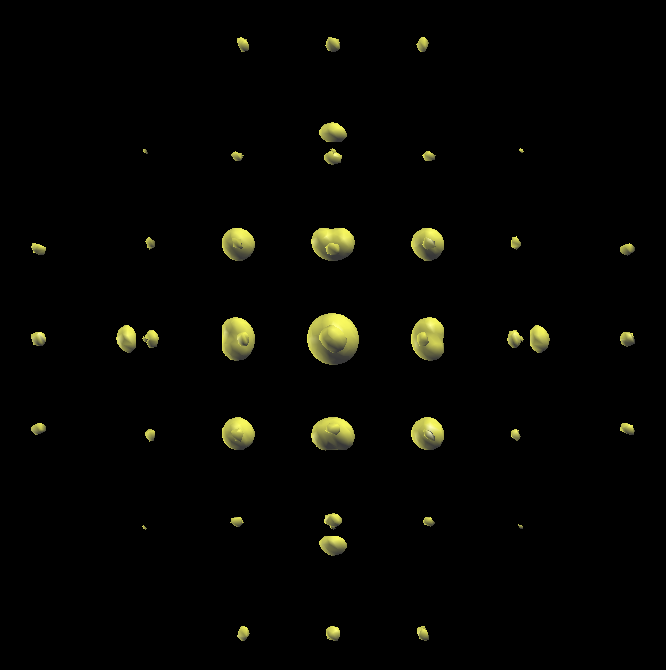}
\includegraphics[width=0.193\textwidth]{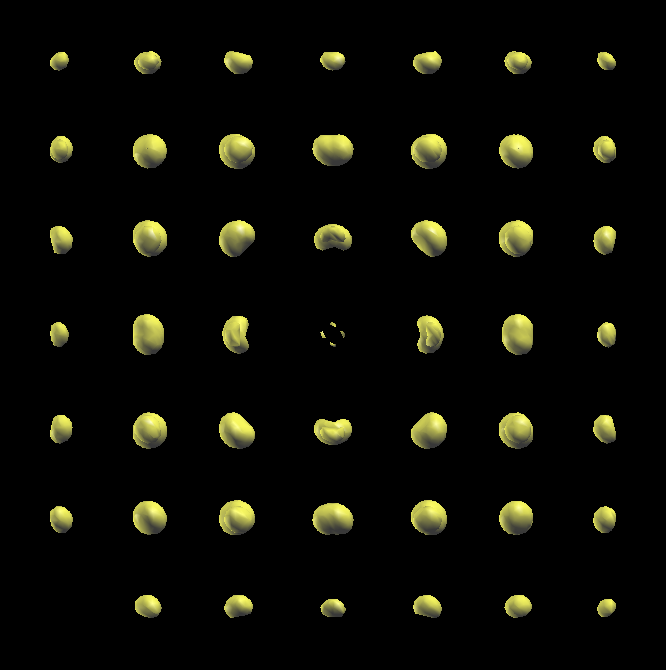}
\end{center}
\caption{Exciton density for the first 5 excitons of LiF at fixed hole position. Upper row, view along the 111 direction. Lower row, view along the 100 direction. The hole position is put nearby a F atom at the center of the picture. The density is averaged over degenerate excitons. All excitons are threefold degenerate, with the exception of the 3rd, which is twofold degenerate. }
\label{fig:exc_wfs}
\end{figure*}

We want now to understand why some transitions are bright, depending on the relative orientation of the pump and the probe pulse. 
A first qualitative explanation can be sought by pursuing the common approach used in literature, based on the a simple hydrogenic model for the excitonic states. To this scope we perform a direct inspection of the excitonic wave-function  and we try to label the excitonic envelop as $s$ or $p$ states.
The overall wave-function symmetry will be the product of the symmetry of the envelop, $A^\lambda_{cv\bfk}$, and the symmetries of the underlying Bloch states, $\psi^*_{c\bfk}(\mathbf{r}_e)$ and $\psi_{v\bfk}(\mathbf{r}_h)$). To this end 
we first notice that for LiF the symmetry group is $O_h$ which has 48 symmetry operations and 10 irreducible representation (for the irreps, see App.~\ref{App:DipSecRules}).
Transitions belonging to the $T_{1u}$ irrep are bright at equilibrium.
The F($2p$) orbitals of the conduction band are associated to the $T_{1u}$ irrep, while the Li($2s$) to the $A_{1g}$ irrep~\cite{Dresselhaus2008}. Since $T_{1u}\times A_{1g}=T_{1u}$, Li($2s$) $\rightarrow$ F($2p$) transitions are dipole allowed.
The first six poles $E_1$...$E_6$ all involve these orbitals. Accordingly we can label exciton-exciton transition involving $E_1$...$E_6$ as \textit{intra}-exciton transitions, which are controlled by the the envelop of the excitonic wave-function.

We thus turn our attention to the description of the envelop.
In Fig.~\ref{fig:exc_wfs} the square modulus of the excitonic wave-functions
\begin{equation}
\Psi_{\lambda}(\mathbf{r}_e,\mathbf{r}_h) = \sum_{cv\bfk} A^\lambda_{cv\bfk} \psi^*_{c\bfk}(\mathbf{r}_e)\psi_{v\bfk}(\mathbf{r}_h)
\end{equation}
is plotted as a function of the electronic coordinate and by fixing the hole coordinate nearby one F atom.
In all the considered excitonic states the wave-functions display non negligible electronic density only on the F atoms. Despite this fact was already observed for the lowest energy exciton~\cite{Rohlfing1998}, and also studied with some details~\cite{Tatewaki1995}, it goes against the intuition of the LiF excitons described in terms of CT excitons, due to the nature of the valence and conduction band structure, which remain used in many model calculations~\cite{Abbamonte2008,ChiCheng2013}. 
In terms of the hydrogenic model for the exciton, since Li($2s$) $\rightarrow$ F($2p$) already fulfill the selection rules, bright excitons at equilibrium are expected to have $l=0$.
Instead in the non equilibrium spectrum, it is only the envelop which determines if exciton-exciton transitions are allowed. Dipole allowed transition should respect $\Delta l=\pm 1$ and $\Delta m_l=0,\pm 1$\cite{Jorger2005}.

Inspecting the plot of the excitonic wave-function,
the two bright excitons $E_1$ and $E_4$ are the only ones where the electronic density sits in the same atom as the hole.
The lowest energy $E_1$ exciton in the literature has been identified as a $1s$ exciton~\cite{Rohlfing1998}. $E_4$ can be identified as the $2s$ exciton. 
For the other dark excitons, the electronic charge is on the nearest neighbor F atoms, as can be seen either from the 111 ($E_2$ and $E_3$) or the 110 ($E_5$ exciton) view, since in all cases there is a negligible electron density in the center of the image.
Accordingly we can tentatively label as $p$-like these dark states. This would explain why these excitons can be seen in the
transient spectrum starting from the $1s$ exciton. However, within this scheme, we are not able to  explain why some are 
bright in the case pump and probe are parallel, while others are bright in the case pump and probe are perpendicular.

A satisfactory analysis of the selection rules governing the transitions among exciton states can be performed only by 
considering the space group representation of crystal lattice \textit{also} for the excitonic envelop.
Let us start with the bright excitons. An $s$ envelop in $O_h$ corresponds to $A_{1g}$ symmetry without any energy lifting~\cite{Dresselhaus2008}, and overall the symmetry of the excitonic wave-function for $E_1$ and $E_4$ is
 $$[s-envelop]\times [F(2s)\rightarrow L(2p)] = A_{1g} \times T_{1u} = T_{1u},$$
which explains why they are bright and threefold degenerate as already observed.
A $2p$ envelop in $O(h)$ is associated to $T_{1u}$ symmetry, again without any energy lifting. The overall symmetry of the underlying excitons is then
\begin{multline*}
[2p-envelop]\times [F(2s)\rightarrow L(2p)] = T_{1u} \times T_{1u} = \\ A_{1g}+E_{g}+T_{1g}+T_{2g},
\end{multline*}
We now need to assign to each exciton the correct representation. We can use the exciton multiplicity, and, again, the knowledge of how a given irrep in $O_h$ can be related to a given angular momentum. The four resulting irreps can all originate by the lifting of spherical states with even angular momentum. However $T_{2g}$ and $E_g$ can originate from states with $l\ge 2$, while $T_{1g}$ and $A_g$ from states with
$l\ge 3$~\cite{Dresselhaus2008}. Thus we expect that the two lowest energy excitons should come from $T_{2g}$ and $E_g$ (i.e. we assume lower angular momentum implies lower energy) and looking at the exciton multiplicity we can assign $E_2[T_{2g}]$ and $E_3[E_{g}]$. The ordering is in agreement with the well known energy splitting of $d$ orbitals in $O_h$ symmetry, i.e. $E_2[T_{2g}]<E_3[E_{g}]$. Since $E_6$ is non degenerate, $E_6[A_{1g}]$, and we tentatively assign $E_5[T_{1g}]$.

We can now use this analysis to predict which exciton-exciton transitions are allowed. First of all we notice that, out of the 10 irreps operations, only the states belonging to the irreps $A_{1g}$, $E_{g}$, $T_{1g}$, and $T_{2g}$ are bright (see again App.~\ref{App:DipSecRules}).
This are exactly the 4 irreps that originates from the $2p$ excitons, and means that all four dark excitons are expected to be visible in the TrAbs spectrum. Moreover the relative polarization of the pump and probe pulses can alternatively select different excitons.
More specifically, the allowed bright representations need to transform as the corresponding quadratic function of the
coordinates consistently with the polarization choice of the pump and probe pulses.  So, when the pump and probe are both parallel to the $x$ axis we see that $A_{1g}$ and $E_{g}$ are expected to be detected. Instead, for perpendicular pump and probe, $T_{2g}$ and $T_{1g}$ are seen. Indeed we see, from the middle panel of Fig.~\ref{fig:abs1} that $E_3[E_g]$ and $E_6[A_{1g}]$ are bright, while $E_2[T_{2g}]$ and $E_5[T_{1g}]$ are bright in the lower panel. This confirms our previous analysis. Finally we observe that $T_{1u} \rightarrow T_{1u}$ transitions are dipole forbidden. Starting from the exited $1s$ state injected by the pump laser pulse (i.e the state in the $T_{1u}$ multiplet with polarization parallel to the pump pulse polarization, we call it $T^x_{1u}$ partner~\cite{Dresselhaus2008}), the transitions towards the other two degenerate states ($T^y_{1u}$ and $T^z_{1u}$) of the multiplet are dipole forbidden.

\subsubsection{Strong pumping regime}

\begin{figure}[t]
	\begin{center}
		\includegraphics[width=0.47\textwidth]{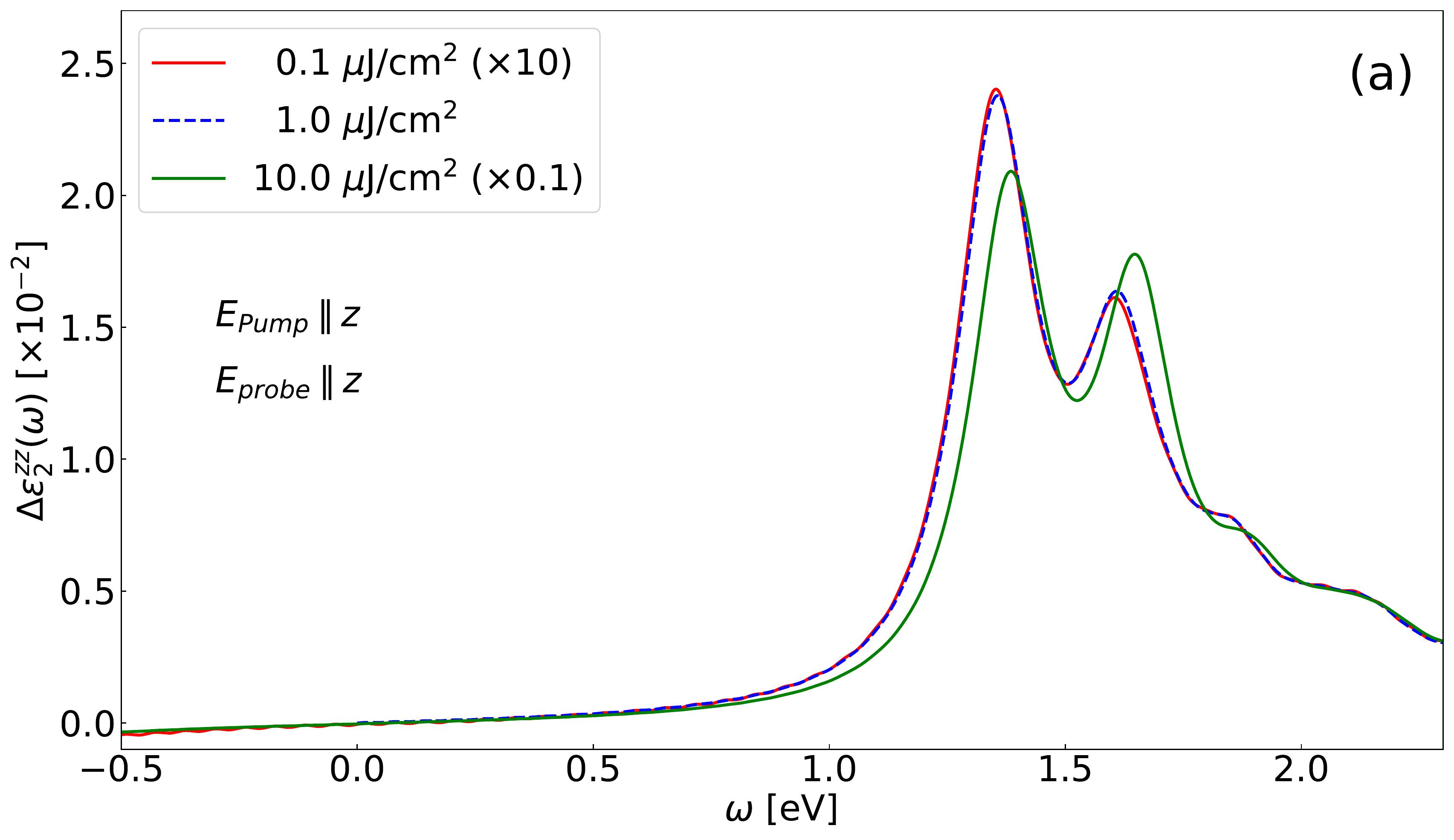}
		\includegraphics[width=0.47\textwidth]{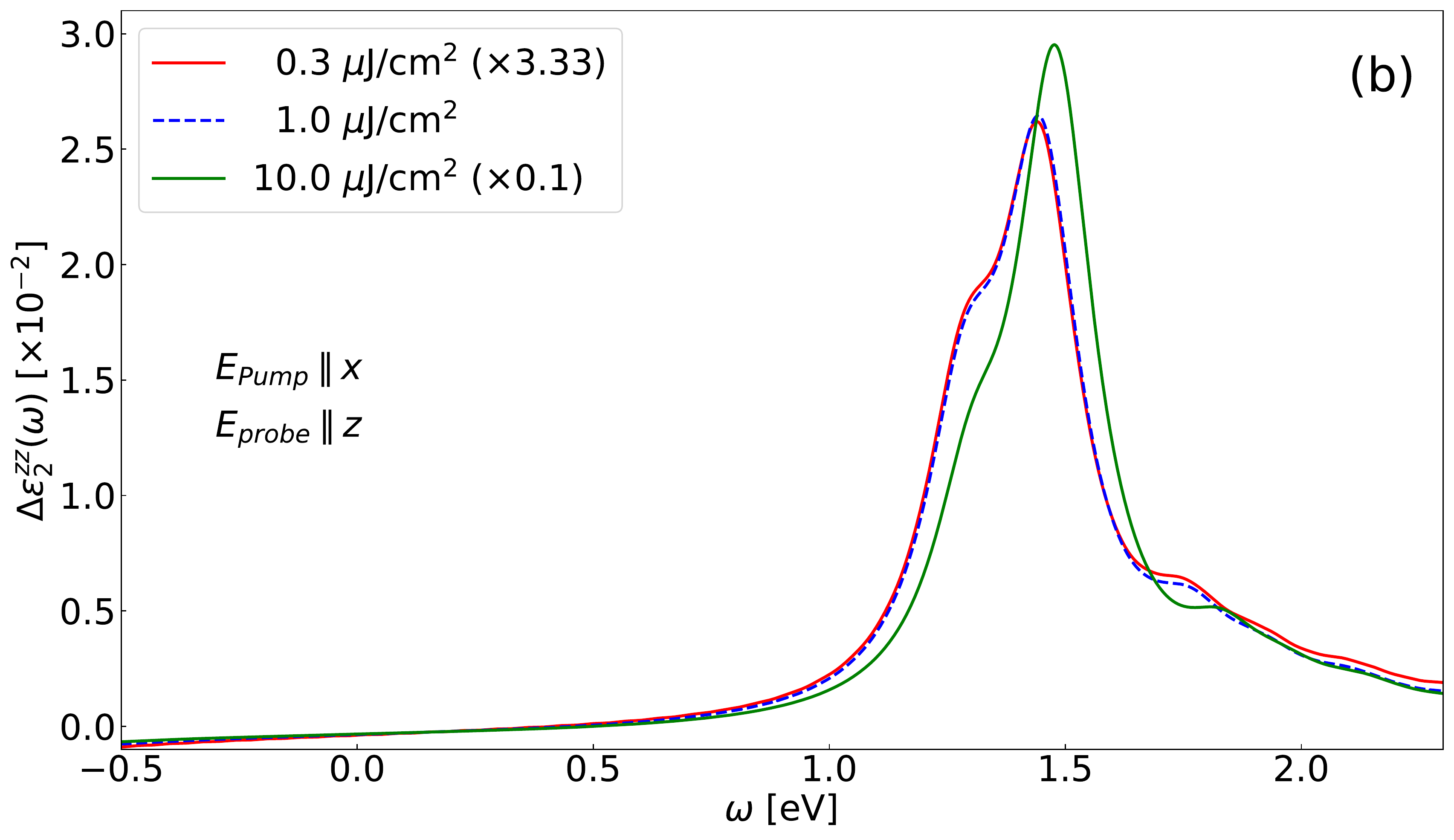}
	\end{center}
	\caption{Transient Absorption defined as in Fig.~1 of the main text as a function of the pump fluence. The reported values for the laser fluence refer to the field inside the sample, i.e. they should be corrected taking into account the fraction of the laser pulse which is reflected when comparing with nominal experimental fluences. Compared to Fig.~\ref{fig:abs1} a larger broadening parameter is used here.}
	\label{fig:abs2}
\end{figure}

We conclude the discussion on LiF by showing how the shape of the TrAbs spectrum changes as a function of the pump fluence. Results are shown in Fig.~\ref{fig:abs2}. The exciton Fermi golden rule approach is not able to reproduce the evolution of the spectrum in this regime, and only the real time propagation scheme can be used. Increasing the pump fluence, and accordingly the initial exciton density, the energy required for the inter-exciton transition is blue-shifted for the first four dark excitons explored in the manuscript. The intensity of the transition $E_1 \rightarrow E_3$ is reduced, while the one for the transition $E_1 \rightarrow E_6$ is enhanced (parallel configuration); the intensity for the transition $E_1 \rightarrow E_2$ is reduced, while the on of the transition $E_1 \rightarrow E_4$ is enhanced (perpendicular configuration).
Inspecting the numerical simulations (and also from the experimental and theoretical literature on TrAbs in the resonant probe regime) we know that the energy position of the $E_1$ peak is blue-shifted due to many-body effects~\cite{Sangalli2016,Trovatello2022}.
For the considered fluences, the injected exciton density is roughly linear with the pump fluence. With a fluence of 1 $\mu$J/cm$^3$ we obtain an exciton density of 1.7 10$^{-4}$ exc./$\Omega$, with $\Omega$ the unit cell volume. This corresponds to a total density of $\approx 10^{19}$ exc./cm$^{3}$. Given the very large binding energy and the very small  exciton radius of $E_1$ in LiF ($r_{exc}\approx 5$ \AA), we expect the critical exciton Mott transition density ($n_{Mott}\sim r_{exc}^{-3}$) to be well above $10^{20}$ exc./cm$^{3}$.~\cite{Snoke2008}
The overall blue-shift of the $E_1 \rightarrow E_I $ transitions implies that the higher energy peaks are shifted even more. Moreover the relative changes in the intensity of the peaks, imply that, not only the excitons shift, but also that there are changes in the excitonic wave-functions which renormalize the expectation value of the intra-exciton dipole matrix elements.

\subsection{Monolayer hexagonal BN}

The structure of electron-hole excitations in monolayer hBN(m-hBN) is well know from theory and experiments~\cite{galvani2016excitons}. In particular, recent measurements probed for the first time the lowest excitonic state of single layer hBN deposited on a substrate \cite{rousseau2021monolayer}.
The electronic excitation of m-hBN can be classified according to the representations of the D$_{3h}$ point group (see also Sec.~\ref{hBN_group}).
In panel $(a)$ of Fig.~\ref{P_and_P_bn} we report the computed optical absorption~\cite{eps2d} of monolayer hBN with in plane polarization. The exciton classification for the first 5 excitonic peaks. $E_1[E']$, $E_2[E']$, $E_3[A'_2]$, $E_4[A'_1]$, and $E_5[E']$, is here taken from Ref.~\onlinecite{galvani2016excitons}.
Among the different group representations, only the two-dimensional irrep $E'$ is optically active for in plane polarization~\cite{Dresselhaus2008}, while excitons belonging to the $A''_2$ irrep are bright for out of plane polarization (i.e. along the z direction). In the present manuscript we focus on the in plane polarization.
{In Ref.~\onlinecite{galvani2016excitons} these excitons are also mapped into the Hydrogen series by comparing \ai\, and tight-banding simulations. A similar analysis is also performed in Ref.~\cite{Zhang2022}.
The lowest exciton $E_1[E']$, can be identified as the $1s$ exciton, and has a very strong oscillator strength. The next three excitons, $E_2[E']$, $E_3[A'_2]$, and $E_4[A'_1]$, originate from $2p$ like state, and at variance with the LiF case, they are not all dark. Finally  the fifth exciton $E_5[E']$ originates from a $2s$ state. The states with $A'_1$ symmetry have been measured in two photon absorption experiments~\cite{Attaccalite2018}, while other states are usually inaccessible because of their too high energy. 

\begin{figure}[t]	
	\begin{center}
        \includegraphics[width=0.45\textwidth]{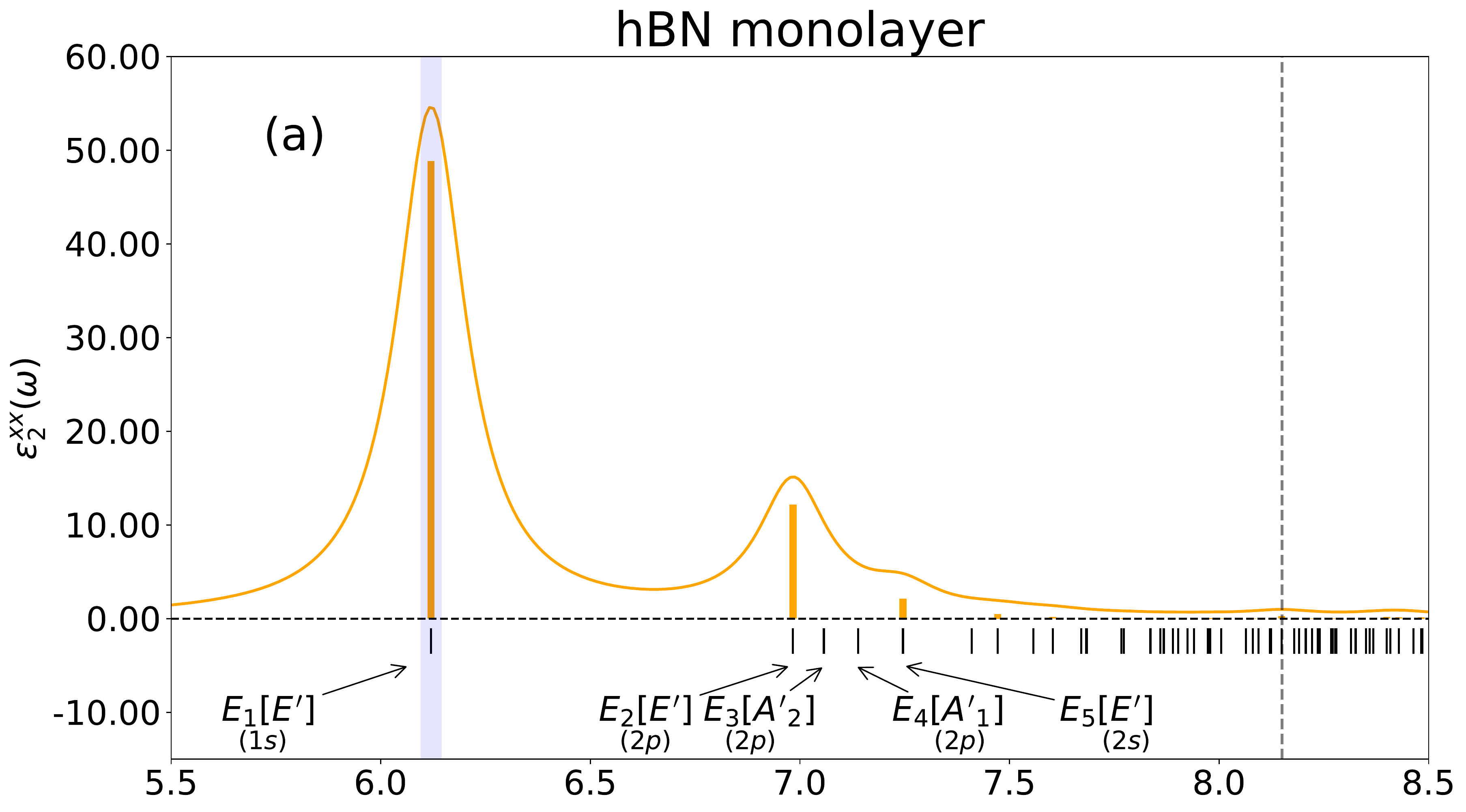}
		\includegraphics[width=0.45\textwidth]{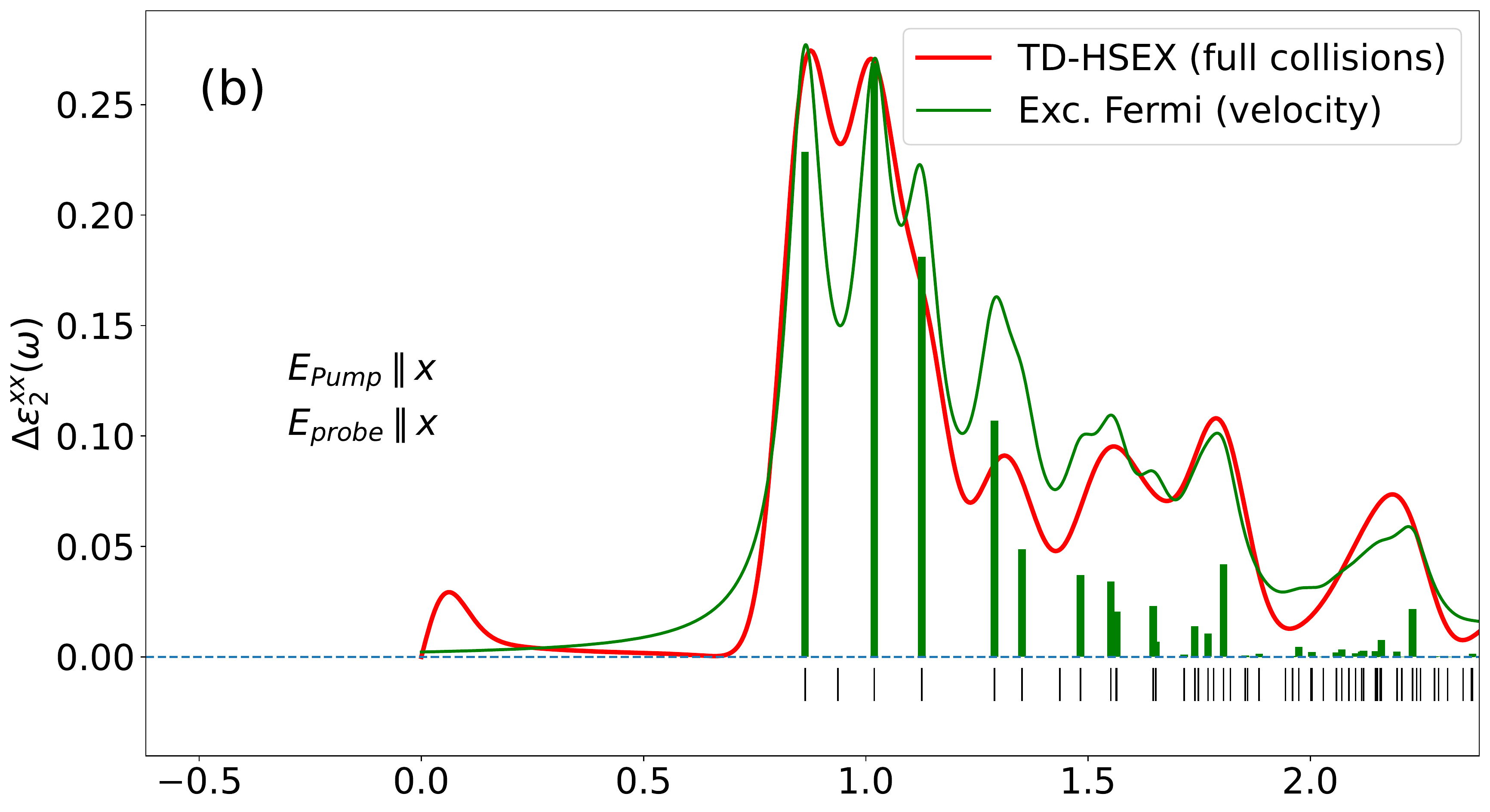}
		\includegraphics[width=0.45\textwidth]{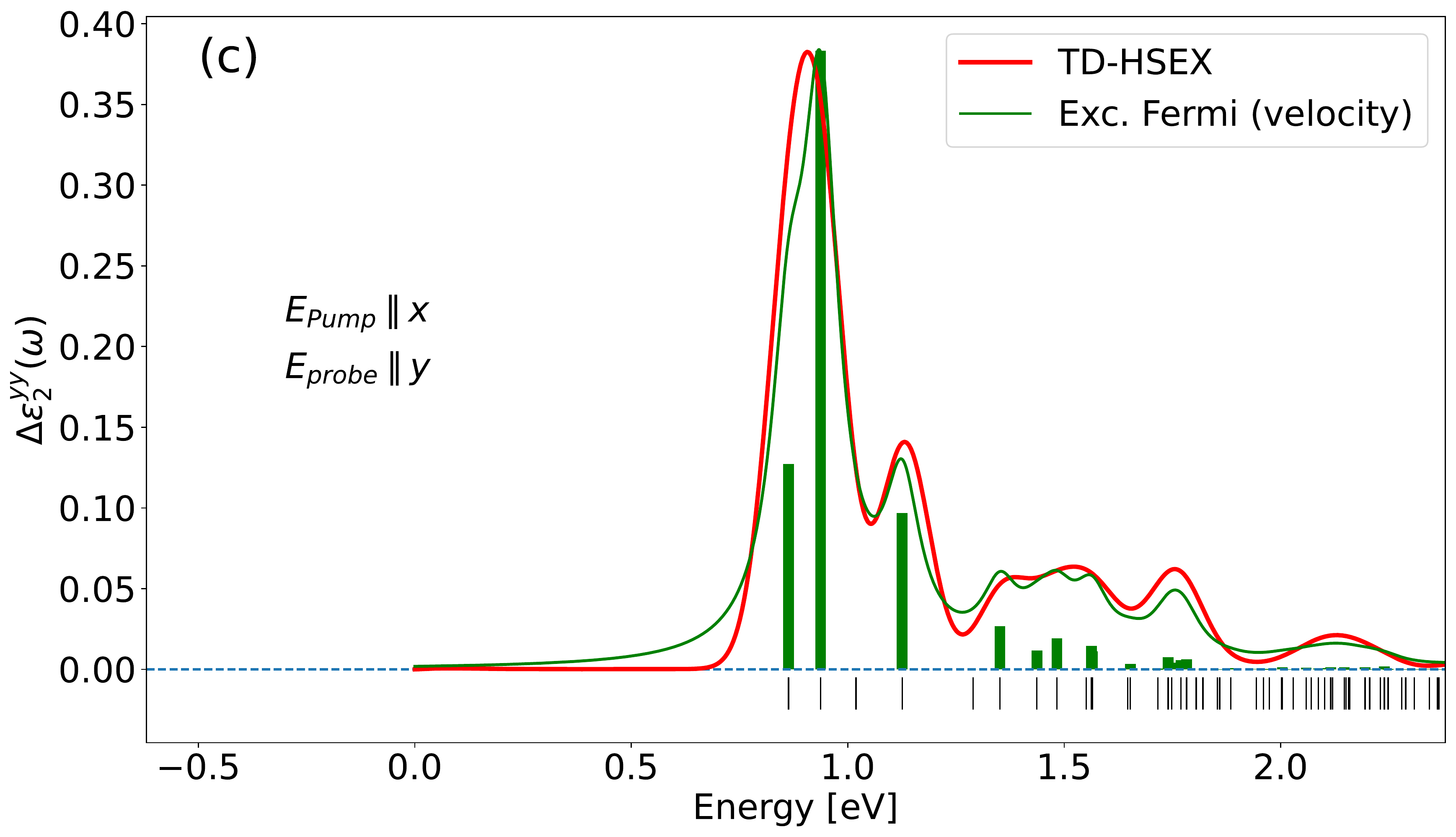}
		\caption{Equilibrium absorption $(a)$ of a hBN monolayer\cite{eps2d} and Pump and probe spectra $(b-c)$ for the monolayer hBN.\label{P_and_P_bn}. Top panel both the pump and probe are in the $x$ direction, while in the bottom 
		panel the pump is along $x$ while the probe $y$. We report in green the excitonic levels at equilibrium and their 
		symmetry for the first four.\cite{galvani2016excitons}}
	    \label{fig:abs1_hBN}
	\end{center}
\end{figure}

Like in the LiF, here we show how pump and probe spectroscopy can be used as alternative technique to the one and two-photon absorption to study excitons in this material. Pump and probe spectroscopy presents a double advantage, firstly the possibility of studying excitons that are dark in linear optics, and secondly  the possibility of studying high-energy excitons that usually are inaccessible to standard laboratory facilities due to the large band gap of hBN~\cite{artus2021ellipsometry}. 
In order to simulate pump and probe spectroscopy in a monolayer hBN, we used a supercell with in plane lattice parameter of $a=2.5~\AA$ and a distance between the periodic replica of $c=30~a.u.$. We constructed the dielectric constant that enters in the BSE and SEX self-energy using 100 bands and 20~Ry cutoff, plus a terminator to speed up convergence on the conduction bands~\cite{Bruneval2008}. In the optical spectra and real-time dynamics we include 2 valence and 2 conduction bands, and a scissor of $3.66~eV$. 
The equations of motion in real-time were propagated for 165~fs using the same algorithm and parameters of the LiF case.
We also restrict ourselves to the case where the probe polarization is in the plane. In panels $(b)-(c)$ of Fig.~\ref{P_and_P_bn} we report the pump and probe spectra for the hBN monolayer in two different cases. In  panel $(b)$ both the pump and the probe are in the $x$ direction, while in panel $(c)$ the pump is along $x$ and the probe along $y$ direction. 
Simlarly to the LiF case, we observe that the TD-HSEX scheme gives poles exactly at the energy differences $E_i-E_1$, with $i$ running over the other excitons.
In the same panels we report also the pump and probe spectra calculated using the Fermi golden rule approach discussed in Sec.~\ref{ssec:FermiGolden} within the velocity gauge. The comparison between the two approaches shows very good agreement, emphasizing that the approximations made in Sec.~\ref{ssec:FermiGolden} introduce only a marginal error in the final spectra. However, at variance with the case of LiF, we have an instability in the real-time numerical simulations for the case of probe parallel to the pump, which gives a small peak at $\omega \approx 0$.

By performing a group theory analysis similar to the LiF case (see App.~\ref{hBN_group}), we find that for parallel pump and probe only excitons with $E'$ and $A'_1$ symmetry can be excited starting from the lowest exciton $1s$ with $E'$ symmetry, while for perpendicular pump and probe field the excitons with $E'$ and $A_2$ symmetry are accessible. This means that bright states are also visible in the transient spectra and in both geometry. This is due to the lower symmetry of the $D_{3h}$ group, and has another remarkable consequence. 
This theoretical prediction are confirmed by our real-time numerical simulations\footnote{Notice that in the real-time calculations we included interactions only between valence and conduction bands.\cite{Attaccalite2011}. We verified that this approximation does not affect the final spectra, see App.~\ref{App:HSEX_self}.}.
Let us remark the consequences of this, and how much hBN is different from LiF.
In the case of LiF the hydrogen atom selection rules $\Delta l = \pm 1$
are not able to predict which transition will be activated by fixing the relative polarization of the pump and the probe. However they are enough to predict that only $1s \rightarrow 2p$ transitions are dipole allowed at low energy.
In hBN instead $\Delta l = \pm 1$ is not fulfilled at all, since the $1s \rightarrow 2s$ transition, i.e.
$E_1[E'] \rightarrow E_5[E']$ transition, is now dipole allowed. 
Moreover the almost degenerate $1s \rightarrow 1s$, i.e. $E^x_1[E'] \rightarrow E^y_1[E']$ energy transitions is now also dipole allowed. This is why a low energy peak appears in the numerical simulations. The exact energy of this peak will depend on the injected exciton density which would be responsible to slightly lift the degeneracy between the two states belonging to $E'$. We did not try to converge it numerically and, moreover, we are not yet able to explain why it appears only in the parallel pump probe polarization configuration.
Still, it is a remarkable result that such transition can be seen in the simulation and a fingerprint that a very low energy the probe pulse is in principle able to spatially rotate the polarization induced by the pump pulse in the xy plane. A mechanism which is not allowed in LiF. 

\section{Conclusions}
\label{sec:conclusions}

In this article, we studied exciton-exciton transitions in transient absorption experiments by performing accurate ab-initio real-time simulations which couples the TD-HSEX scheme with the Berry Phase expression for the polarization.
To interpret the results obtained via this real-time scheme, we also develop a Fermi golden rule approach based on the assumption that the initial state created by the pump laser pulse can be well described as a quasi-equilibrium state with a well defined excitonic population.
In this way, the non-equilibrium response can be analyzed in terms of transitions between different excitons.
Starting from the excitonic wave functions computed with the ab-initio Bethe Salpeter equation, we define the dipole elements between the different excitonic states, and use them to calculate the nonequilibrium response function. This is also corroborated by a detailed group theory analysis of the excitonic states.

The agreement obtained via the formally more rigorous real time propagation schemes shows that for laser pulse intensities compatible with many pump and probe experiments, bound excitons such as those in LiF and hBN behave as well-defined quasi-particles. Moreover it offers a validation of the Fermi Golden Rule approach, which is numerically much cheaper, and potentially interesting for future applications. While the real-time propagation scheme could be employed to study the effect of increasing exciton density as well as relaxation and dissipation mechanism, the Exciton Fermi golden rule approach provides a cheaper approach, where also the signature of finite momentum excitons into TrAbs spectra can be investigated.

Finally these results show that absorption spectroscopy in the exciton-exciton energy range, offers new opportunities to study high-energy excited states not accessible by other spectroscopic techniques. The technique is also a very powerful tool to measure exciton relaxation and exciton dynamics.

\section*{Acknowledgments}
This publication is based upon work from COST Action TUMIEE CA17126, supported by COST (European Cooperation in Science and Technology).
The work was also in part supported by European Union project MaX Materials design at the eXascale H2020-EINFRA-2015-1 (Grant Agreement No. 824143) and by Italian Miur PRIN (Grant No. 20173B72NB and 2020JZ5N9M), and the ANR grant Colibri (Project No. ANR-22-CE30-0027).
C.A. acknowledges A. Saul and K. Boukari for the management of the computer cluster \textit{Rosa}. DS aknowledges useful discussions with Fulvio Paleari.

\appendix

\section{Derivation of the dipole matrix elements}
In this appendix we present the derivation of dipole matrix elements between different excitonic states, both in the equilibrium and non-equilibrium case.

\subsection{Equilibrium case: ground-state to exciton transition}

First we derive Eq.~\eqref{eq:dipole_1part} in the formalism of second quantization. This is a well known result and the derivation is here reported just as a preliminary step towards the derivation of Eq.~\eqref{eq:dipole_exc_complete}. We introduce the creation operator $\ac{n\bfk}$ which creates an electron in the state
$\ket{n\bfk}$ acting on the many-body vacuum state $\ket{0}$. The GS and the valence-conduction
pair are written as
\be\lb{state_sq1}
\ket{g} = \prod_{v\bfk}^{N_v}\ac{v\bfk}\ket{0} \, , \quad
\ket{cv\bfk} = \ac{c\bfk}\ad{v\bfk}\ket{g} \, .
\ee
The (many-body) position operator expressed in this formalism reads
\be\lb{position_sq2}
\dipole = \sum_{nm\bfk} \ac{n\bfk}\ad{m\bfk}r_{nm\bfk} \, , \quad \textrm{where} \quad 
r_{nm\bfk} = \bra{n\bfk}\opr\ket{m\bfk} \, .
\ee
Using the definition of the excitonic state $\bra{\lambda\bfq}$, the matrix element \eqref{eq:dipole_1part} is written as
\begin{eqnarray}
\bra{\lambda\bfq} \dipole \ket{g} & = & \sum_{cv\bfk} A^{\lambda\bfq}_{cv\bfk} \bra{cv\bfk}\dipole\ket{g} \, , \\
\bra{cv\bfk}\dipole\ket{g}  &=& \sum_{nm\bfk}
\bra{g}\ac{v\bfk}\ad{c\bfk}\ac{n\bfk}\ad{m\bfk}\ket{g} r_{nm\bfk} \nn \, ,
\end{eqnarray}
and we have to remember that $c$ is in the conduction sector, $v$ is in the valence sector and the
sum over $n$ and $m$ in unrestricted. Looking at the bracket we see that $m$ has to be in the 
valence sector, otherwise $\ad{m\bfk}\ket{g} = 0$. The bracket is the scalar product of two
\emph{states} that is not zero only if $m=v$ and $n=c$, so we obtain
\be 
\bra{cv\bfk}\dipole\ket{g}  = r_{cv\bfk} \, , 
\ee 
which is the expected result.

The definition of the position operator used here is ill defined in periodic boundary conditions, in particular because the intra-band terms are ill defined. For systems with a gap this is not an issue in the derivation of Eq.~\eqref{eq:dipole_1part} since only matrix elements with $n$ in valence and $m$ in conduction are involved. This problem will appear in the definition of the dipoles for exciton to exciton transitions, which will depend on term with $n=m$.
One possible solution is to move from the length to the velocity gauge via the definition of the current operator $j$, another involves the formal definition of $r=\partial_\bfk$.
We do not explicitly address this issue in the present appendix.

\subsection{Nonequilibrium case: exciton to exciton transitions}\label{App:ExcExcDips}
We now proceed with the derivation of Eq.~\eqref{eq:dipole_exc_complete}.
In the case of an initial state containing one exciton three options can be considered for the final state for $\ket{J}$.
(i) $\ket{F}=\ket{g}$, i.e. the probe stimulates the emission of a photon, back to the ground state.
(ii) $\ket{F}=\ket{\lambda_i\bfq_i+\lambda_f\bfq_f}$ with $E_F=E_0+\omega_{\lambda_i+\lambda_f}(\bfq_i+\bfq_f)$, i.e. the probe stimulates the creation of a second exciton. The energy of the two excitons (or bi-exciton) state can be in general lower then the sum of the two excitonic energies ($\omega_{\lambda_i+\lambda_f}(\bfq_i+\bfq_f)<\omega_{\lambda_i}(\bfq_i)+\omega_{\lambda_f}(\bfq_f)$) , due to the extra bi-exciton binding energy.\cite{schafer2013semiconductor}
(iii) $\ket{F}=\ket{\lambda_f\bfq_f}$ with $E_F=E_0+\omega_{\lambda_f}(\bfq_f)$, i.e. the initial exciton is further excited into a different excitonic state. Due to the optical nature of the transition, we must have $\bfq_i=\bfq_f$
Omitting the momentum indexes and using $\omega_\alpha=\omega_{\lambda_i+\lambda_f} $ the expression for the response function has the form
\begin{eqnarray}
\lb{Chi_def2}
\x_{AB}(\w)
&=& \frac{2}{V}\sum_{f}\frac{A_{\l f} B_{f\l}^j}{E_f-E_\l-\w-i\G} \, , \nonumber \\
&=& \frac{2}{V}\frac{A_{\l g} B_{g\l}}{-\omega_\l-\w-i\G}
    + \frac{2}{V}\sum_{\alpha}\frac{A_{\l\a} B_{\a\l}}{\omega_\a-\omega_\l-\w-i\G}  \nonumber \\ 
    &&\ \  + \frac{2}{V} \sum_{\l'}\frac{A_{\l\l'} B_{\l'\l}}{\omega_\l'-\omega_\l-\w-i\G}
     \, ,
\end{eqnarray}
While in the first two cases (i) and (ii) peaks are expected in the energy range defined by $\omega_\lambda(\bfq)$, we here focus into this latter case (iii), i.e. the last term in the equation, where peaks are expected at much lower energies. 

We use the same approach of the previous section to derive a formula for the dipole matrix elements
for exciton to exciton transitions. In this case we have to deal with a matrix element of the form
\be\lb{dipole_exc}
\bra{cv\bfk}\dipole\ket{c'v'\bfk} \, ,
\ee
where $c,c'$ are in the conduction sector, while $v,v'$ are in the valence one.
Plugging the expressions for excitonic states and position operator provides
\be\lb{dipole_exc2}
\sum_{nm\bfk}
\bra{g}\ac{v\bfk}\ad{c\bfk}\ac{n\bfk}\ad{m\bfk}\ac{c'\bfk}\ad{v'\bfk}\ket{g} r_{nm\bfk} \, .
\ee
We split the sum over $n$ and $m$ in the valence and conduction sector. We observe that if $m$ 
is in conduction we must have $m=c'$, while if $n$ is in conduction we must have $n=c$.
Accordingly we have four terms
\begin{align}
\sum_{n,n \in occ,\bfk}
&\bra{g}\ac{v\bfk}\ad{c\bfk}\ac{n\bfk}\ad{m\bfk}\ac{c'\bfk}\ad{v'\bfk}\ket{g} r_{nm\bfk} + \nn \\
+\sum_{m \in occ,\bfk}
&\bra{g}\ac{v\bfk}\ad{m\bfk}\ac{c'\bfk}\ad{v'\bfk}\ket{g} r_{cm\bfk} + \nn \\
+\sum_{n \in occ,\bfk}
&\bra{g}\ac{v\bfk}\ad{c\bfk}\ac{n\bfk}\ad{v'\bfk}\ket{g} r_{nc'\bfk} + \nn \\
+&\bra{g}\ac{v\bfk}\ad{v'\bfk}\ket{g} r_{cc'\bfk} \, . \nn
\end{align}
Now we observe that the second term can be non zero only if $m=c'$ and the third if $n=c$, which
are both non allowed conditions because $m$ and $n$ are in the valence sector.  So the expression
reduces to
\be 
\sum_{n,m \in occ,\bfk}
\bra{g}\ac{v\bfk}\ad{c\bfk}\ac{n\bfk}\ad{m\bfk}\ac{c'\bfk}\ad{v'\bfk}\ket{g} r_{nm\bfk} +
\d_{vv'}r_{cc'\bfk} \, . \nn
\ee
We analyze the first bracket and we move $\ac{v\bfk}$ toward the right. This gives
\be 
\d_{vm}\bra{g}\ad{c\bfk}\ac{n\bfk}\ac{c'\bfk}\ad{v'\bfk}\ket{g} + 
\d_{vv'}\bra{g}\ad{c\bfk}\ac{n\bfk}\ad{m\bfk}\ac{c'\bfk}\ket{g} \, , \nn
\ee 
now we observe that, in order to be non zero the ket and bra states have to be equal, so we match 
the indexes in the only way and we obtain
\be 
-\d_{vm}\d_{nv'}\d_{cc'} + \d_{vv'}\d_{nm}\d_{cc'} \, . \nn
\ee 
Putting everything together we obtain
\be\lb{dipole_exc3}
\bra{cv\bfk}\dipole\ket{c'v'\bfk} = -\d_{cc'}r_{v'v\bfk} + 
\d_{vv'}\d_{cc'}\sum_{n\in occ}r_{nn\bfk} + \d_{vv'}r_{cc'\bfk}  \, .
\ee 
Finally, we can insert this formula in the dipole matrix element of two excitonic states. This provides
\begin{align}     
\bra{\l}\dipole\ket{\l'} = \sum_{cv\bfk}\sum_{c'v'\bfk}A^{\l*}_{cv\bfk}A^{\l'}_{c'v'\bfk}
\bra{cv\bfk}\dipole\ket{c'v'\bfk} \nn \, , 
\end{align}
and using \eqref{dipole_exc3} we find
\begin{align} 
\bra{\l}\dipole\ket{\l'} = 
&-\sum_{cvv'\bfk}A^{\l*}_{cv\bfk}A^{\l'}_{cv'\bfk}r_{v'v\bfk}
+\sum_{cvc'\bfk}A^{\l*}_{cv\bfk}A^{\l'}_{c'v\bfk}r_{cc'\bfk} + \nn \\
&+\sum_{cv\bfk}A^{\l*}_{cv\bfk}A^{\l'}_{cv\bfk}\sum_{n\in occ}r_{nn\bfk} \nn \\
= 
&\sum_{v,c\neq c',\bfk}A^{\l*}_{cv\bfk}A^{\l'}_{c'v\bfk}r_{cc'\bfk}
-\sum_{c, v\neq v',\bfk}A^{\l*}_{cv\bfk}A^{\l'}_{cv'\bfk}r_{v'v\bfk} \nn \\
&+\sum_{v,c,\bfk}A^{\l*}_{cv\bfk}A^{\l'}_{cv\bfk}(r_{cc\bfk}-r_{vv\bfk} +\sum_{n\in occ}r_{nn\bfk})
\, . \lb{dipole_exc_complete_app}
\end{align}

A very similar expressions was derived in Ref.~\cite{bechstedt}, Eq.~(B8), to model non linear optics in the independent particles case. A similar expression is also derived also in Ref.~\cite{taghizadeh2018gauge} (B3), a part from the last term.

\section{Dipole allowed transitions by symmetry analysis} \label{App:DipSecRules}

\subsection{LiF bulk and the $O(h)$ symmetry group}

The point symmetry group of LiF is $O_h$ which has 48 symmetry operations and 10 irreducible representation (irreps), 4 of dimension 3 ($T_{1u},T_{2u},T_{1g},T_{2g}$), 2 of dimension 2 ($E_u, E_g$), and 4 of dimension 1 ($A_{1u},A_{2u},A_{1g},A_{2g}$).


For the case of equilibrium absorption, the initial state corresponds to the ground state $| \Psi_I \rangle=| \Psi_0 \rangle$, which is symmetric, thus $O_I=A_{1g}$ irrep. The dipole operator instead $\dipole$ belongs to the $T_{1u}$ irrep. Then, given the irreducible transformation $O_F$ of the final state $| \Psi_f \rangle$, the integral is different from zero if the product of the symmetries gives $A_1g$, i.e.
\begin{equation}
A_{1g} \in A_{1g} \times T_{1u} \times O_F \, , 
\end{equation}
to this end we need to look into the product table for the $O_h$ group \cite{Dresselhaus2008}.
From the table we see that $A_{1g} \times T_{1u}=T_{1u}$ and the condition $T_{1u}\times O_f= A_{1g}$ is verified only for
\begin{equation}
{O_f=T_{1u}} \, , 
\end{equation}
$T_{1u}$ has dimension 3, and $\Psi_f$ is threefold degenerate.

We move to the nonequilibrium case where the pump generates a bright excitonic state, thus $O_I=T_{1u}$. In this case
we need to perform the same exercise as before, looking for the states for which:
\begin{equation}
A_{1g} \in T_{1u} \times T_{1u} \times O_F \, .
\end{equation}
Since $T_{1u}\times T_{1u}=A_{1g}+E_{g}+T_{1g}+T_{2g}$, and $A_{1g} \in O_F \times O_F$
We have that any
\begin{equation}
O_F= A_{1g},E_{g},T_{1g},T_{2g} \, , 
\end{equation}
is a possible solution. This already shows that the inter-exciton transition spectrum will show excitons which are dark when measuring absorption from equilibrium. Bright to bright transitions are not possible.
A further push in the identification of the bright excitations can be achieved by analyzing also the relative polarization of the pump and probe pulses. Indeed, if both the pulses are parallel to (say) the 
$z$-direction, the associated irreps have to transform as the corresponding quadratic form of the coordinates, \emph{i.e.} as $z^2$ in this specific case. Then, the character table of $O_h$ unambiguously identifies the first bright
excitation (which is double degenerate) as $E_g$ and the second one as $A_{1g}$. Instead for transverse
polarization (with the probe parallel to the $x$ axis), the allowed irreps have to transform as the bilinear product $xz$. This requirement is satisfied by both $T_{2g}$ which transform as the quadratic form $(xz,yz,xy)$ and
$T_{1g}$ which transform as the rotation $(R_x,R_y,R_z)$. So in this case this arguments allows us to
identifies the two bright states as $T_{1g}$ and $T_{2g}$ excitons. 

\subsection{hBN monolayer and the $D(3h)$ symmetry group}
\label{hBN_group}
For the case of the hBN monolayer the point group is $D_{3h}$ which has 6 irreps: $E'$ and $E''$, of dimension 2, 
and $A'_1$, $A'_2$, $A''_1$,$A''_2$, all of dimension 1. The ground state is symmetric and belongs to the $A_1$ 
irrep.

The $x$ and $y$ components of the dipole operator, which are the relevant elements for the study of the in-plane excitation belong to the $E'$ irrep, while the $z$ one belongs to the $A''_2$.
Following the same reasoning done for LiF, the in-plane optically active excitons belong to the $E'$ irrep and are twofold degenerate, while the out-of plane optically active excitons belong to $A''_2$.

Moving to the non-equilibrium absorption the initial configuration belongs to $E'$ so the coupling
with an in-plane dipole gives rise to $(E'\times E')=(A'_1+A'_2+E')$, i.e. it spans all the possible irreps in the \textit{primed sector} of the point group. So, any exciton in this sector can be reached 
measuring absorption from an optical excited state, independently of the fact that the exciton is bright or dark at equilibrium.

Also in this case the polarization based analysis allows us to discriminate the representation to
which the bright states belong. Indeed, for pulses both parallel to (say) the $x$-axis the excited
states has to transform has the $x^2$ quadratic form, so only $E'$ (with a double degeneracy) and $A'_1$ 
(non degenerate) are allowed. On the contrary for transverse in-plane polarization the excited state
transform as the bilinear $xy$, so only $E'$ and $A'_2$ (that transforms as the $R_z$ rotation)
are allowed.
 
\section{Selection of degenerate excitonic state}\label{App:DegenerateExc}

In presence of degenerate bright excitons, we select, in the degenerate excitonic space $U_d^i$, the specific linear combination of states which gives an excitonic transition dipole along the direction of the pump laser pulse $\efield_0$.
To this end we define a ``dipoles matrix'' and a ``directions matrix'':
\begin{eqnarray}
D_{\alpha\lambda}=\frac{\mu^\alpha_{0\lambda}(\mathbf{0})}{|\boldsymbol{\mu}_{0\lambda}(\mathbf{0})|} \\
E_{\alpha n}=\frac{\mathcal{E}^\alpha_n}{|\efield_n|}
\end{eqnarray}
These two matrices are $n_{deg} \times n_{deg}$, depending on the size of the degenerate space. For LiF $n_{deg}=3$, and the transition dipoles of the 3 states spans the whole space. For hBN $n_{deg}=2$, and the transition dipoles spans the xy plane.
Here $\lambda$ are the indexes of the exciton in $U_d^i$, while $n$ runs from $0$ to $n_{deg}-1$. $\efield_n$, with $n>0$, are directions orthogonal to $\efield_0$ that belong to the space spanned by the transition dipoles.
Thanks to these two matricides we can define the rotation matrix
\begin{equation}
C_{\lambda n}=\sum_{\alpha} D^t_{\lambda\alpha} E_{\alpha n}.
\end{equation}
$C_{\lambda 0}$ is used to construct the excitonic states
\begin{equation}
 | \lambda_n \rangle =	\sum_{\lambda} C_{\lambda 0} | \lambda \rangle.
\end{equation}
The states $| \lambda_n \rangle$ span the same space spanned by the random vectors $| \lambda \rangle$, but now the initial state $| \lambda_i \rangle = | \lambda_0 \rangle$ has polarization parallel to $E_{\alpha 0}$.

\section{Intraband term: position vs velocity dipoles}

Intra-band dipoles $r_{nn\bfk}$ are ill defined in the length gauge. They can instead  be accounted within the velocity gauge and thus shifting to the (one body) velocity dipole matrix elements $\mathbf{v}^{1b}_{nm\bfk}$. In the linear regime starting from a non-equilibrium state, position and velocity dipoles can be related by the  expression:
\begin{equation}
\mathbf{v}^{1b}_{nm\bfk}=i\,\mathbf{r}_{nm\bfk}\Delta\epsilon_{nm\bfk} + \delta_{n,m} \partial_\bfk \epsilon_{n\bfk} .
\end{equation}
The above equation can be found in many works in the literature~\cite{Virk2007} for the case $\Delta\epsilon_{nm\bfk}>0$, while the expression for the terms with $n=m$ can be obtained by the expression for the energies in the $\bfk\cdot \mathbf{v}$ model~\cite{Luppi2010} in presence of non-degenerate bands:
${\epsilon_{n\bfk+\bfq}\simeq \epsilon_{n\bfk} + \bfq\cdot \mathbf{v}_{nn\bfk} }$.

Notice that such equation also shows that the physics of intraband transitions enters differently in the two gauges. Indeed, while for $\Delta\epsilon_{nm\bfk}>0$ the two dipoles carry the same information, for $n=m$ the velocity dipoles carry extra information which cannot be obtained from the length dipoles alone. However, while the velocity dipoles capture the physics of intra-band transitions, their use within the velocity gauge has two main drawbacks: (i) sum rules are easily broken in numerical implementations, and (ii) beyond the independent-particles approximation the definition of the velocity operator depends on the Hamiltonian~\cite{sangalli2017optical}.

\section{Approximated HSEX self-energy}} \label{App:HSEX_self}
The $\Delta \Sigma^{HSEX}[\rho(t)]$ appearing in Eq.~\ref{eq:hmb} is calculated in the Yambo code as:\cite{Attaccalite2011}
\begin{equation}
\label{eq:SEX_expression}
\Delta \Sigma^{\text{HSEX}}_{mm' \bfk}(t) = \sum_{n,n'\bfq} M_{\substack{mm'\bfq \\ nn'\bfk}} \cdot \Delta \rho_{n,n'\mathbf{k-q}}(t).
\end{equation}
where the matrix elements of $M$ are defined as the sum of two terms:
\begin{eqnarray}
M^{H}_{\substack{mm'\bfq \\ nn'\bfk}} =  \sum_{\bfG} \rho^\bfG_{mm'\bfk}(\mathbf{0}) \[\rho^\bfG_{nn' \bfk-\bfq}(\mathbf{0})\]^* v_{\bfG}(\bfq) , \label{eq:H_kernel} \\
M^{SEX}_{\substack{mm'\bfq \\ nn'\bfk}} =  \sum_{\bfG,\bfG'} \rho^{\bfG'}_{mn\bfk} (\bfq) \[\rho^\bfG_{m'n' \bfk}(\bfq)\]^* W_{\bfG,\bfG'}(\bfq),  \label{eq:SEX_kernel}
\end{eqnarray}
where 
\begin{equation}
\rho^{}_{mn\bfk} (\bfq,\mathbf{G}) = \int  \varphi^*_{m \bfk}( \mathbf r) \varphi_{n\mathbf{k-q}}( \mathbf r)  e^{i(\mathbf G+\mathbf q) \mathbf r}.
\end{equation}
$v$ and $W$ are the bare and the screened Coulomb interaction already introduced in the main text.\cite{aryasetiawan1998gw} In this work we considered only matrix elements of $M$ involving both $m,n$ in the valence (conduction) when  $m',n'$ are both in the conduction (valence) band. Instead, we set to zero all elements where either $m,m'$ or $n,n'$ are both in the conduction (valence) band. This approximation strongly reduces the computational cost, is exact for the linear response, and we verified that does not introduce any relevant change in the pump and probe spectra reported in the main text.

\bibliography{biblio.bib}

\begin{thebibliography}{79}
\bibitem [{eps()}]{eps2d}%
  \BibitemOpen
  \href@noop {} {}\bibinfo {note} {For a 2D material we define the effective
  imaginary part of the dielectric constant as the 2D polarizability per
  surface unit $\alpha^{2D}$ divided by an effecting distance along the $z$
  direction. For hBN we used $L_{eff}=3.25$~\r{A}, $\epsilon_2^{eff}(\omega) =
  Im[\alpha^{2D}(\omega)] /L_{eff}$}\BibitemShut {NoStop}%
\end{thebibliography}

\end{document}